\documentclass[usenatbib]{mnras}
\usepackage{graphicx}
\usepackage{amsmath}
\usepackage{amssymb}
\usepackage{color}
\usepackage{url}

\usepackage{threeparttable}

\newcommand\lsim{\mathrel{\rlap{\lower4pt\hbox{\hskip1pt$\sim$}}
        \raise1pt\hbox{$<$}}}
\newcommand\gsim{\mathrel{\rlap{\lower4pt\hbox{\hskip1pt$\sim$}}
        \raise1pt\hbox{$>$}}}
\newcommand{\lya}{\ifmmode\mathrm{Ly}\alpha\else{}Ly$\alpha$\fi}
\newcommand{\lyb}{\ifmmode\mathrm{Ly}\beta\else{}Ly$\beta$\fi}
\newcommand{\igm}{\ifmmode\mathrm{IGM}\else{}IGM\fi}
\newcommand{\lae}{\ifmmode\mathrm{LAE}\else{}LAE\fi}
\newcommand{\h}{\ifmmode\mathrm{H}\else{}H\fi}
\newcommand{\hi}{\ifmmode\mathrm{H\,{\scriptscriptstyle I}}\else{}H\,{\scriptsize I}\fi}
\newcommand{\hii}{\ifmmode\mathrm{H\,{\scriptscriptstyle II}}\else{}H\,{\scriptsize II}\fi}
\newcommand{\cmb}{\ifmmode\mathrm{CMB}\else{}CMB\fi}
\newcommand{\qso}{\ifmmode\mathrm{QSO}\else{}QSO\fi}
\newcommand{\eor}{\ifmmode\mathrm{EoR}\else{}EoR\fi}
\newcommand{\heii}{\ifmmode\mathrm{He\,{\scriptscriptstyle II}}\else{}He\,{\scriptsize II}\fi}
\newcommand{\heiii}{\ifmmode\mathrm{He\,{\scriptscriptstyle III}}\else{}He\,{\scriptsize III}\fi}
\newcommand{\ciii}{\ifmmode\mathrm{C\,{\scriptscriptstyle III]}}\else{}C\,{\scriptsize III]}\fi}
\newcommand{\oiii}{\ifmmode\mathrm{O\,{\scriptscriptstyle III}}\else{}O\,{\scriptsize III}\fi}
\newcommand{\aliii}{\ifmmode\mathrm{Al\,{\scriptscriptstyle III}}\else{}Al\,{\scriptsize III}\fi}
\newcommand{\mgii}{\ifmmode\mathrm{Mg\,{\scriptscriptstyle II}}\else{}Mg\,{\scriptsize II}\fi}
\newcommand{\fe}{\ifmmode\mathrm{Fe}\else{}Fe\fi}
\newcommand{\nv}{\ifmmode\mathrm{N\,{\scriptscriptstyle V}}\else{}N\,{\scriptsize V}\fi}
\newcommand{\niv}{\ifmmode\mathrm{N\,{\scriptscriptstyle IV]}}\else{}N\,{\scriptsize IV]}\fi}
\newcommand{\cii}{\ifmmode\mathrm{C\,{\scriptscriptstyle II}}\else{}C\,{\scriptsize II}\fi}
\newcommand{\civ}{\ifmmode\mathrm{C\,{\scriptscriptstyle IV}}\else{}C\,{\scriptsize IV}\fi}
\newcommand{\siv}{\ifmmode\mathrm{Si\,{\scriptscriptstyle IV}}\else{}Si\,{\scriptsize IV}\fi}
\newcommand{\siii}{\ifmmode\mathrm{Si\,{\scriptscriptstyle II}}\else{}Si\,{\scriptsize II}\fi}
\newcommand{\siiii}{\ifmmode\mathrm{Si\,{\scriptscriptstyle III]}}\else{}Si\,{\scriptsize III]}\fi}
\newcommand{\ovi}{\ifmmode\mathrm{O\,{\scriptscriptstyle VI}}\else{}O\,{\scriptsize VI}\fi}
\newcommand{\sioiv}{\ifmmode\mathrm{Si\,{\scriptscriptstyle IV}\,\plus O\,{\scriptscriptstyle IV]}}\else{}Si\,{\scriptsize IV}\,+O\,{\scriptsize IV]}\fi}

\newcommand{\avenf}{$\bar{x}_{\rm HI}$}

\newcommand{\intermediateHII}{\textsc{\small Intermediate \hii{}}}

\newcommand{\cmfst}{\textsc{\small 21CMFAST}}

\pdfoutput=1
\title[Damping wing constraints from two $z\gtrsim7$ QSOs]{IGM damping wing constraints on reionisation from covariance reconstruction of two $z\gtrsim7$ QSOs}
\author[B. Greig et al.] {Bradley~Greig$^{1,2}$\thanks{E-mail:~greigb@unimelb.edu.au}, Andrei~Mesinger$^{3}$, Frederick B.~Davies$^{4}$, Feige Wang$^{5}$\thanks{NHFP Hubble Fellow}, \newauthor
Jinyi Yang$^{5}$\thanks{Strittmatter Fellow}, Joseph F.~Hennawi$^{6,7}$\\
$^1$School of Physics, University of Melbourne, Parkville, VIC 3010, Australia \\
$^2$ARC Centre of Excellence for All-Sky Astrophysics in 3 Dimensions (ASTRO 3D) \\
$^3$Scuola Normale Superiore, Piazza dei Cavalieri 7, I-56126 Pisa, Italy\\
$^4$Max-Planck-Institut f{\"u}r Astronomie, K{\"o}nigstuhl 17, 69117 Heidelberg, Germany \\
$^5$Steward Observatory, University of Arizona, 933 N Cherry Ave, Tucson, AZ 85721, USA \\
$^6$Department of Physics, Broida Hall, University of California, Santa Barbara, CA 93106-9530, USA \\
$^7$Leiden Observatory, Leiden University, P.O. Box 9513, 2300 RA Leiden, the Netherlands
}

\begin{document}
\maketitle \begin{abstract}
\noindent
Bright, high redshift ($z>6$) QSOs are powerful probes of the ionisation state of the intervening intergalactic medium (IGM).
The detection of Ly$\alpha$ damping wing absorption imprinted in the spectrum of high-z QSOs can provide strong constraints on the epoch of reionisation (EoR). In this work, we perform an independent \lya{} damping wing analysis of two known $z>7$ QSOs; DESJ0252-0503 at $z=7.00$ (Wang et al.) and J1007+2115 at $z=7.51$ (Yang et al.). For this, we utilise our existing Bayesian framework which simultaneously accounts for uncertainties in: (i) the intrinsic \lya{} emission profile (reconstructed from a covariance matrix of measured emission lines; extended in this work to include \nv{}) and (ii) the distribution of ionised (\hii{}) regions within the IGM using a $1.6^3$ Gpc$^3$ reionisation simulation. This approach is complementary to that used in the aforementioned works as it focuses solely redward of \lya{} ($1218 < \lambda < 1230$\AA) making it more robust to modelling uncertainties while also using a different methodology for (i) and (ii). We find, for a fiducial EoR morphology, $\bar{x}_{\hi{}} = 0.64\substack{+0.19 \\ -0.23}$ (68 per cent) at $z=7$ and $\bar{x}_{\hi{}} = 0.27\substack{+0.21 \\ -0.17}$ at $z=7.51$ consistent within $1\sigma$ to the previous works above, though both are slightly lower in amplitude. Following the inclusion of \nv{} into our reconstruction pipeline, we perform a reanalysis of ULASJ1120+0641 at $z=7.09$ (Mortlock et al.) and ULASJ1342+0928 at $z=7.54$ (Ba{\~n}ados et al.) finding $\bar{x}_{\hi{}} = 0.44\substack{+0.23 \\ -0.24}$ at $z=7.09$ and $\bar{x}_{\hi{}} = 0.31\substack{+0.18 \\ -0.19}$ at $z=7.54$. Finally, we combine the QSO damping wing constraints for all four $z\gtrsim7$ QSOs to obtain a single, unified constraint of $\bar{x}_{\hi{}} = 0.49\substack{+0.11 \\ -0.11}$ at $z=7.29$.
\end{abstract} 

\section{Introduction}

Following the recombination of protons and electrons at the surface of last scattering, the Universe is permeated by a dense, cold fog of neutral hydrogen. This fog is only lifted when the cumulative ionising radiation from stars, galaxies and QSOs exceeds the number of neutral Hydrogen atoms plus its recombinations rate. This baryonic phase change is referred to as the epoch of reionisation (EoR). Reionisation proceeds in a patchy manner: individual ionised (\hii{}) regions around the first structures merge with their neighbours before percolating and eventually ionising the entire intergalactic medium (IGM). The timing and duration of the EoR can loosely be inferred through integral measurements such as the Thomson scattering of cosmic microwave background (CMB) photons off the liberated electrons (measured as an optical depth, $\tau_{\rm e}$; e.g. \citealt{Planck:2018}). 

Recent observational evidence appears to point towards the tail-end of reionisation occurring relatively late. For example, the observed 110 $h^{-1}$~Mpc Gunn-Peterson down to $z=5.5$ \citep{Becker:2015}, the large scatter in the \lya{} effective optical depth at $z\sim5.5$ \citep{Eilers:2018,Yang:2020a, Bosman:2021a} and the \citet{Bosman:2018} \lya{} forest data that when interpreted in a forward-modelled Bayesian framework implies reionization completed at $z\lesssim5.6$ (\citealt{Qin:2021}, see also \citealt{Kulkarni:2019,Keating:2020a,Keating:2020b,Nasir:2020,Choudhury:2021}).

Direct constraints on reionisation via the absorption of \lya{} photons by lingering neutral regions at $z\gtrsim5$ are problematic due to the amplitude of the scattering cross-section of \lya{} photons in the resonant core. Here, the IGM is in photo-ionisation equilibrium whereby the density of the residual neutral hydrogen ($x_{\hi} \gtrsim 10^{-4} - 10^{-5}$) is sufficient to saturate transmission \citep{Fan:2006p4005}. Thus order unity fluctuations in the neutral fraction during reionisation are difficult to distinguish from fluctuations in the ultraviolet background, density or temperature, post reionisation. Alternatively, a more robust probe of the IGM neutral fraction is available through the \lya{} damping wing (e.g. \citealt{Rybicki1979,MiraldaEscude:1998p1041}). The absorption cross-section of the extended Lorentzian wings in the \lya{} line profile are several orders of magnitude below that of the line centre, making them ideal for studying the order unity fluctuations in neutral fraction during reionisation.

Constraints on the IGM neutral fraction from the damping wing have been successfully employed using both galaxies and QSOs. For galaxies, the imprint is only measurable after averaging over large statistical samples \citep[e.g.][]{Mesinger:2015p1584,Mason:2018,Hoag:2019,Mason:2019}. QSOs on the other hand, being orders of magnitude brighter, enable the damping wing to be detectable from a single spectrum \citep[e.g.][]{Mesinger:2007,Mortlock:2011p1049,Bolton:2011p1063,Bosman:2015,Greig:2017,Banados:2018,Davies:2018,Greig:2019,Dominika:2020,Wang:2020,Yang:2020}. However, care must be taken to properly characterise the individual QSO intrinsic emission, together with the corresponding uncertainties.

There are several approaches in the literature specifically tailored toward reconstructing the intrinsic QSO emission profile. One of the most well studied approaches is through principal component analysis (PCA). Here, spectral features of the QSO continuum are deconstructed into principal component vectors. These principal components are separated redward and blueward of $\lambda=1280$\AA\ with either a linear \citep{Davies:2018a,Bosman:2021} or neural network based \citep{Dominika:2020} mapping used to reconstruct the blue-side from the measured red-side components. These mappings are learned from $\sim13,000$ moderate-$z$ QSOs. Alternatively, both \citet{Fathivavsari:2020} and \citet{Liu:2021} have employed deep neural networks (i.e. deep learning). In the former, a network was trained on $\sim18,000$ high signal to noise (S/N) QSOs using spectral information from the \civ{}, \siv{} and \ciii{} broad emission lines to predict \lya{}. In the latter, the network was trained on high resolution Hubble Space Telescope spectra at $z\sim0.2$ to predict the entire $1020{\rm \AA} \leq \lambda \leq 1600{\rm \AA}$ range from spectral information provided between $1216{\rm \AA} \leq \lambda \leq 1600{\rm \AA}$. Finally, \citet{Reiman:2020} employ conditional neural spline flows to provide a fully probabilistic reconstruction of the $1190{\rm \AA} \leq \lambda \leq 1290{\rm \AA}$ region from redward information over $1290{\rm \AA} \leq \lambda \leq 2900{\rm \AA}$.

In this work, we use the Bayesian reconstruction pipeline developed in \citet{Greig:2017a}. Here, the \lya{} profile is reconstructed using a covariance matrix of correlations between the \lya{} line and other high ionisation emission lines (e.g. \civ{}, \siv{} and \ciii{}). This assumes that all emission line profiles are modelled by a single, or two component Gaussian profile characterised entirely by its width, height and velocity offset. Further, that the correlations themselves are fully encapsulated by an N-dimensional normal distribution. This Gaussian assumption appears justified based on its performance relative to our training set data.

Importantly, our full damping wing analysis pipeline combines the reconstructed profile with a distribution of synthetic IGM damping wings extracted from a large-scale EoR simulation \citep{Mesinger:2016p6167}. This step accounts for the statistical uncertainties that arise from the fact we are only considering single sightlines (QSO spectra) to place constraints on the EoR. A similar approach is adopted by \citet{Davies:2018} whose method is applied in \citet{Wang:2020} and \citet{Yang:2020}. However, unlike these works we consider only the damping wing imprint redward of \lya{}, avoiding the uncertainties associated with modelling the QSO host environment. For all other works, damping wing constraints are extracted using the simplistic \citet{MiraldaEscude:1998p1041} damping wing model.

Specifically, we apply our analysis framework to the two recently discovered $z\gtrsim7$ QSOs: DESJ0252-0503 at $z=7.00$ (hereafter J0252; \citealt{Yang:2019,Wang:2020}) and J1007+2115 at $z=7.51$ (hereafter J1007; \citealt{Yang:2020}). Our analysis (discussed in detail below), notably differs from the previous ones, thus this work serves as an independent and complementary verification of the inferred EoR constraints. For reference, these authors recover constraints on the IGM neutral fraction of $\bar{x}_{\hi{}} = 0.70\substack{+0.20 \\ -0.23}$ at $z=7.0$ from J0252 and $\bar{x}_{\hi{}} = 0.39\substack{+0.22 \\ -0.13}$ at $z=7.51$ from J1007 (both at 68 per cent confidence).

This work is structure as follows. In Section~\ref{sec:Method} we provide a brief description of the observational data, before outlining our analysis pipeline. In Section~\ref{sec:Results} we provide the main results and in Section~\ref{sec:Discussion} our discussion. In Section~\ref{sec:Conclusion} we conclude with our closing remarks. Unless stated otherwise, we quote all quantities in co-moving units and adopt the cosmological parameters:  ($\Omega_\Lambda$, $\Omega_{\rm M}$, $\Omega_b$, $n$, $\sigma_8$, $H_0$) = (0.69, 0.31, 0.048, 0.97, 0.81, 68 km s$^{-1}$ Mpc$^{-1}$), consistent with recent results from the Planck mission \citep{PlanckCollaboration:2016p7780}.

\section{Method} \label{sec:Method}

In Section~\ref{sec:data}, we first provide a brief summary of the observational data for J0252 and J1007 used in this work. In Section~\ref{sec:Reconstruction} we summarise the covariance matrix reconstruction pipeline introduced by \citet{Greig:2017a} including the modifications for this work and in Section~\ref{sec:Damping} we outline the simulation procedure for extracting the synthetic damping wing profiles. Finally, in Section~\ref{sec:JointFitting} we summarise the joint fitting procedure used to obtain constraints on the IGM neutral fraction.

\subsection{Observational data} \label{sec:data}

\subsubsection{J1007+2115}

In this work, we make use of the high quality combined spectrum outlined in \citet{Yang:2020}. This combines a 5.5 hr (on-source) Gemini Near-Infrared Spectrograph (GNIRS) observation (R $\sim620$) with a 2.2 hr (on-source) observation with Keck/Near-Infrared Echellette Spectromenter (NIRES; R $\sim2700$) using inverse-variance weighting. The systemic redshift for this QSO, $z=7.5149\pm0.0004$, is obtained from [\cii{}] emission using the Atacama Large Millimeter/submillimeter Array (ALMA).

\subsubsection{DESJ0252-0503}

For J0252, we use the combined spectrum outlined in \citet{Wang:2020}. This combines a high quality 4.8 hr on-source near-infrared observation (R $\sim2700$) from Keck/NIRES with earlier Gemini Multi-Object Spectrogram observations (combined R $\sim1300$). The systemic redshift for this QSO, $z=7.000\pm0.001$, is obtained from [\cii{}] emission using the IRAM Northern Extended Millimeter Array (NOEMA).

\subsection{Covariance reconstruction of the intrinsic Ly$\alpha$ profile} \label{sec:Reconstruction}

Previously, in \citet{Greig:2017a} we developed a method to reconstruct the intrinsic \lya{} line profile using a measured covariance matrix of correlations amongst emission line parameters describing \lya{} and other readily measurable high ionisation lines (namely, \civ{}, \sioiv{} and \ciii{}). This covariance matrix was obtained from a carefully cultivated training set\footnote{QSOs were removed in cases where the \lya{} line profile could not be well characterised by our fitting procedure, most commonly caused by excessive absorption at or near \lya{}, see Appendix C in \citealt{Greig:2017a} for further details.} of 1673 moderate-$z$ ($2.08 < z < 2.5$), high signal to noise (S/N $>15$) QSOs from SDSS-III (BOSS) DR12 \citep{Dawson:2013p5160,Alam:2015p5162}.

Each line profile was modelled as a Gaussian, fully described by its peak height, width and velocity offset from systemic. For \lya{} and \civ{} we found a strong preference for a two component Gaussian, containing both a broad and narrow line profile. In total, this amounts to an $18\times18$ covariance matrix describing all emission line correlations.

Importantly, in this work, we extend this model to additionally include the correlations amongst the \nv{} line, resulting in an expanded $21\times21$ covariance matrix of emission line parameters. Previously, we neglected the \nv{} line as it did not strongly correlate with \lya{} and because the rest-frame line centre ($\lambda_{\nv} = 1240.81$\AA) fell outside the region over which we fit for the damping wing imprint ($\lambda = 1218 - 1230$\AA). However, the \nv{} line profile can be blue-shifted closer to this region while also be being sufficiently broad to contribute flux into this wavelength window. Thus, in this work we include \nv{} to account for this. In Appendix~\ref{sec:Covariance_wNV} we present our updated correlation matrix including \nv{}, while in Appendix~\ref{sec:reanalysis} we reanalyse the previous $z\gtrsim7$ QSOs for which we have performed a damping wing analysis for, following the inclusion of \nv{} into our updated reconstruction pipeline.

Following the extension of our covariance matrix to include \nv{}, our reconstruction of the intrinsic \lya{} line profile proceeds as follows:
\begin{itemize}
\item We fit the observed rest-frame QSO spectrum over the range $\lambda =1275-2300$\AA\ using the systemic redshifts\footnote{Note, our training set of BOSS QSOs uses the SDSS-III pipeline redshift to convert to rest-frame rather than a [\cii{}] redshift as used for the observed $z\gtrsim7$ QSOs in this work. While this can lead to biases in the recovered line profile blueshifts, this is likely sub-dominant relative to the scatter in \lya{} reconstruction profiles and synthetic damping wings following our analysis pipeline, see \citet{Greig:2019} for further details.} outlined in Section~\ref{sec:data}.
\item We perform a joint fit of a power-law continuum component plus the Gaussian profiles for the high-ionisation lines \sioiv{}, \civ{} (2-component) and \ciii{} emission line profiles. To improve the overall fit, we also fit for other known emission line profiles as well as simultaneously fitting a variable number of absorption lines each modelled as a single Gaussian profile. This is shown for J1007 and J0252 in Figures~\ref{fig:FitJ1007} and~\ref{fig:FitJ0252}, respectively.
\item Using the fits to the \civ{}, \sioiv{} and \ciii{} emission line components (amplitude, width and velocity offset), we collapse the 21-dimensional (Gaussian distributed) covariance matrix into a 9 dimensional estimate of the joint \lya{} and \nv{} intrinsic emission line profile (two Gaussian components for \lya{} and one for \nv{}).
\item With this 9 dimensional covariance matrix, we then draw joint intrinsic \lya{} and \nv{} profiles from this distribution. We additionally apply a flux prior over the range, $1250 < \lambda < 1275$\AA, to ensure our reconstructed profiles match the flux level of the observed spectrum over this range.
\end{itemize}

\subsection{Modelling the IGM damping wing during reionisation} \label{sec:Damping}

In order to quantify a putative imprint of the neutral IGM in the QSO spectrum we must create synthetic IGM damping wing profiles. Specifically, our synthetic profiles are computed using the Evolution of 21-cm Structure (EOS; \citealt{Mesinger:2016p6167})\footnote{http://homepage.sns.it/mesinger/EOS.html} 2016 simulations. These are large volume (side length of 1.6~Gpc over 1024 voxels) semi-numerical reionisation simulations that include state-of-the-art sub-grid prescriptions for inhomogeneous recombinations and photo-heating suppression of star-formation. In this simulation suite there are three different EoR morphologies, characterised by different star-formation efficiencies for the typical halos hosting star-forming galaxies. 

In \citet{Greig:2019}, we explored the impact of these different EoR morphologies on the recovered constraints on the IGM neutral fraction from the damping wing, finding only weak evidence for a morphology dependence\footnote{Quantitatively, the constraints on the IGM neutral fraction owing to different EoR morphologies was found to be $\pm0.05$ relative to the \intermediateHII\ model. In contrast, the typical errors on our damping wing constraints are $\pm0.15-0.20$.}. As such, in this work, we shall only consider one EoR morphology, the \intermediateHII\ model, which is the middle of the three EoR simulations. This model is characterised by an EoR driven by galaxies residing in $M_h \gsim 10^9 M_\odot$ haloes, consistent with recent results from forward-modelling the \lya{} forest opacity distributions and ultra-violet (UV) galaxy luminosity functions \citep[LFs;][]{Qin:2021}.

Out of this simulation we extract a total of $10^5$ synthetic IGM damping wing profiles. For the $10^4$ largest identified halos (corresponding to roughly $6\times10^{11} < M_h < 3\times10^{12}$ $M_\odot$) we draw 10 randomly oriented sightlines originating from the central host halo. When constructing the synthetic damping wing profiles, we exclude the cumulative contribution from all encountered \hi{} patches inside the measured QSO near-zone for either object. Implicitly, we are assuming that most of the neutral gas inside the proximity zone (at least down to Lyman-limit systems) will have been ionised by the host QSO. This is motivated by the fact we do not observe any evidence for high-column density systems within the proximity zone. For J0252, the estimated near-zone size is $\sim13.6$ comoving Mpc (1.7 physical Mpc) while for J1007 it is $\sim12.6$ comoving Mpc (1.5 physical Mpc). Then, to have these synthetic damping wing profiles vary with \igm{} neutral fraction, we sample the ionisation fields obtained from different redshift snapshots from the simulation\footnote{This assumes that the halo locations do not significantly change due to bulk motions across the different redshift snapshots. Relative to the resolution of the simulation (cell size $\sim$1.6 Mpc) this is a reasonable assumption.}. In total, we have 24 unique values of the \igm{} neutral fraction.

\subsection{Joint fitting to obtain IGM neutral fraction constraints} \label{sec:JointFitting}

Having outlined the individual procedures to obtain both the reconstructed intrinsic \lya{} profile and the synthetic damping wing profiles, we now summarise the joint pipeline used to extract constraints on the IGM neutral fraction from the damping wing imprint. In summary the steps are as follows:
\begin{enumerate}
\item We draw $\sim10^5$ estimates of the intrinsic QSO profile from the combined reconstruction of the \lya{} and \nv{} line profiles as described in Section~\ref{sec:Reconstruction}.
\item For each intrinsic profile, we multiply it by the 10$^5$ synthetic damping wing opacities obtained from Section~\ref{sec:Damping}. This results in $\sim10^{10}$ mock spectra for each \avenf\ snapshot from the EoR simulation.
\item Each of these $\sim10^{10}$ mock spectra are then compared to the observed QSO spectrum over $1218$\AA\ $ < \lambda < 1230$\AA\ (consistent with \citealt{Greig:2017,Greig:2019}). For each mock spectra we calculate its $\chi^{2}$ relative to observed flux and the error spectrum. Additionally, where appropriate, we mask out any features (e.g. absorption) over the $1218$\AA\ $ < \lambda < 1230$\AA\ range that may erroneously bias the resultant fit.
\item The resulting likelihood, averaged (i.e. marginalised) over all $\sim10^{10}$ mock spectra, is then assigned to the \avenf\ of the current simulation snapshot.
\item We then repeat steps (ii)--(iv) for all available \avenf\ snapshots (of which there are 24) to obtain a final 1D probability distribution function (PDF) of \avenf\ for the observed QSO.
\end{enumerate}

\section{Results} \label{sec:Results}

\subsection{Reconstruction of J1007}

\begin{figure*} 
	\begin{center}
	  \includegraphics[trim = 0.5cm 0.6cm 0cm 0.5cm, scale = 0.45]{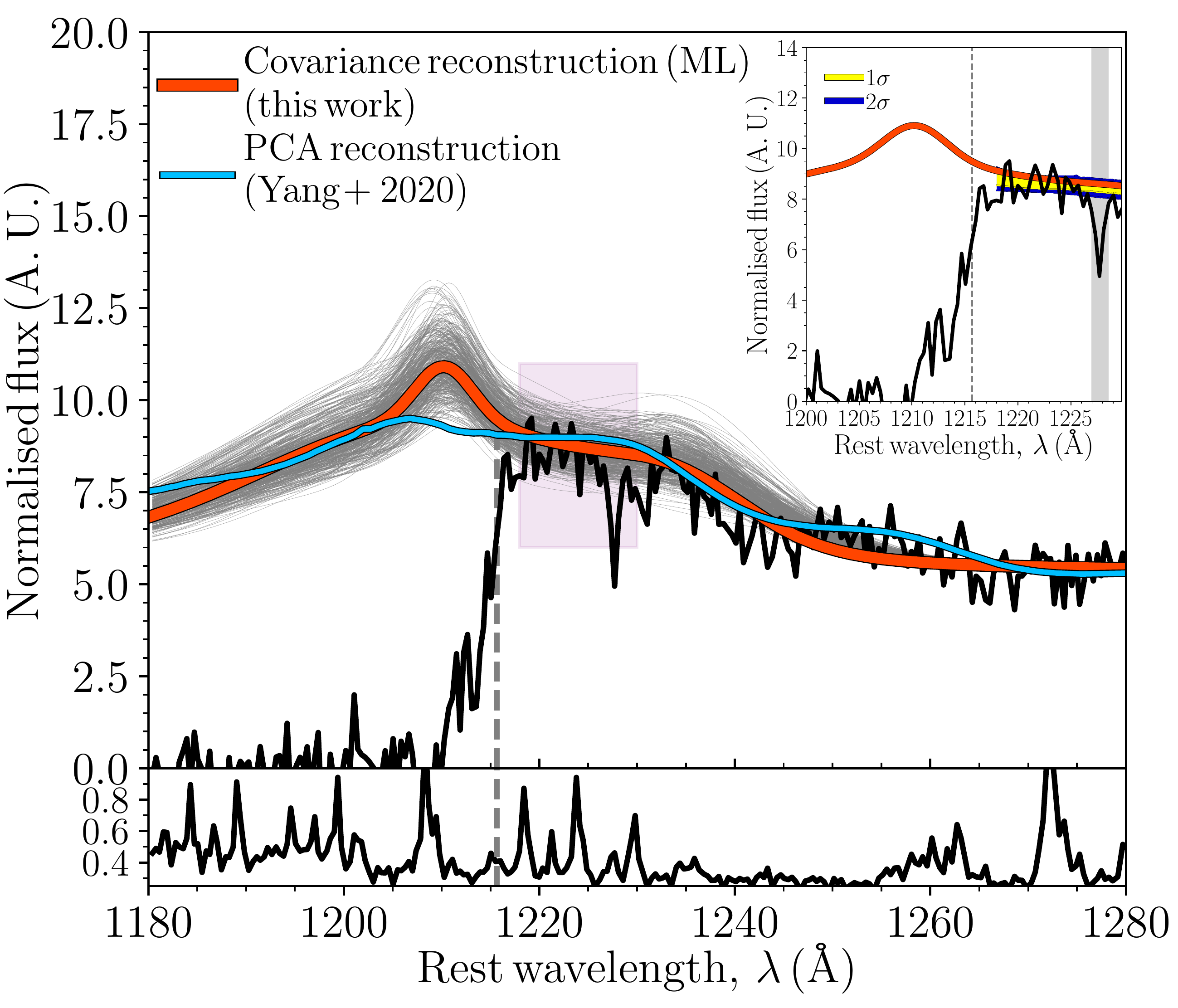}
	\end{center}
\caption[]{The reconstructed maximum likelihood \lya{} emission line profile for J1007 at $z=7.51$ (red curve; shown for visualisation purposes only) and a subsample of 300 \lya{} line profiles (thin grey curves) randomly drawn from the full posterior distribution of reconstructed profiles (see Section~\ref{sec:Reconstruction}; i.e.\ the analysis pipeline samples the full posterior rather than just the maximum likelihood profile). The black curve (and associated error below) corresponds to the combined GNIRS and NIRES spectrum and the blue line corresponds to the PCA reconstructed profile from \citet{Yang:2020} and the grey dashed line denotes rest-frame \lya{}. The purple shaded region corresponds to the region over which we fit for the IGM damping wing imprint, corresponding to the range $\lambda=1218-1230$\AA.}
\label{fig:ReconJ1007}
\end{figure*}

In Figure~\ref{fig:ReconJ1007}, we provide our reconstructed intrinsic QSO profile for J1007. The best-fit (maximum likelihood; ML) reconstruction profile is given by the red curve while the thin grey curves highlight a random subsample of 300 reconstruction profiles drawn from the full posterior distribution. In comparison, the blue curve corresponds to the PCA reconstructed profile from \citet{Yang:2020} using the method outlined in \citet{Davies:2018a}. Note, our reconstruction pipeline does not predict \siii{} ($\lambda\sim1263$), which is included in the PCA pipeline, explaining the different flux levels observed just blueward of 1260\AA. Additionally, for reference in Figure~\ref{fig:FitJ1007}, we provide the MCMC fit of J1007 that is used to obtain the high ionisation line parameters used in our reconstruction pipeline.

Qualitatively speaking, from Figure~\ref{fig:ReconJ1007} we recover noticeably different reconstruction profiles across the two different pipelines; covariance matrix in this work and the PCA approach from \citet{Davies:2018a}. We find a clear preference for a narrow component, peaked blueward of \lya{} at $\lambda\sim1210$\AA\ along with a considerable broad component. In the PCA approach, no clear peak in \lya{} is found, though there is a small peak closer to $\lambda\sim1205$\AA. Both pipelines however do recover similarly broad \nv{} line components. Interestingly though, the velocity offset of the small amplitude \lya{} peak and the \nv{} peak from the PCA approach are similarly blueshifted relative to the \lya{} and \nv{} peaks extracted from our covariance matrix approach indicating that the correlations in the velocity offsets between \lya{} and \nv{} are consistent between the two pipelines. The similarities and differences between the two pipelines (covariance matrix and PCA) have been discussed in detail in \citet{Greig:2019}, however we will return to this discussion in Section~\ref{sec:Discussion}. Nevertheless, despite the different observable profile features, statistically speaking the two different reconstruction methods are consistent with one another when comparing the full distribution of reconstructed profiles\footnote{A more rigorous quantitative analysis between the two approaches is beyond the scope of this paper. While for both methods it is trivial to produce uncertainties for the reconstruction profiles as a function of wavelength, this is not an accurate comparison as the reconstructed profiles themselves are correlated across wavelength bins. Thus, what is actually required is a comparison between the covariance matrices of reconstructed profiles from the two pipelines. As such, we defer such a comparison to future work.}.

Across the region for which we fit the damping wing ($1218$\AA\ $ < \lambda < 1230$\AA), the amplitudes of the reconstructed profiles from the two pipelines are relatively similar, with our covariance matrix approach preferring a slightly lower amplitude flux. However, our reconstructed narrow \lya{} line profile does enter at the edge of our fitting region (near $\lambda\sim1218$\AA), where our reconstructed profile becomes larger in amplitude.

In the inset of Figure~\ref{fig:ReconJ1007}, we present the confidence intervals following our joint fitting procedure (Section~\ref{sec:JointFitting}) with the yellow (blue) shaded regions corresponding to the 68 (95) percentiles. The vertical grey band demarcates a notable absorption feature which we mask from our damping wing fitting interval. As the ML (red) curve is only marginally offset (larger) than the yellow/blue regions this suggests that there is evidence of only a weak IGM damping wing imprint. That is, the intrinsic \lya{} line profile does not require significant attenuation by a neutral (or partially) neutral IGM to match the observed QSO spectrum.

\subsection{Reconstruction of J0252}

\begin{figure*} 
	\begin{center}
	  \includegraphics[trim = 0.5cm 0.6cm 0cm 0.5cm, scale = 0.45]{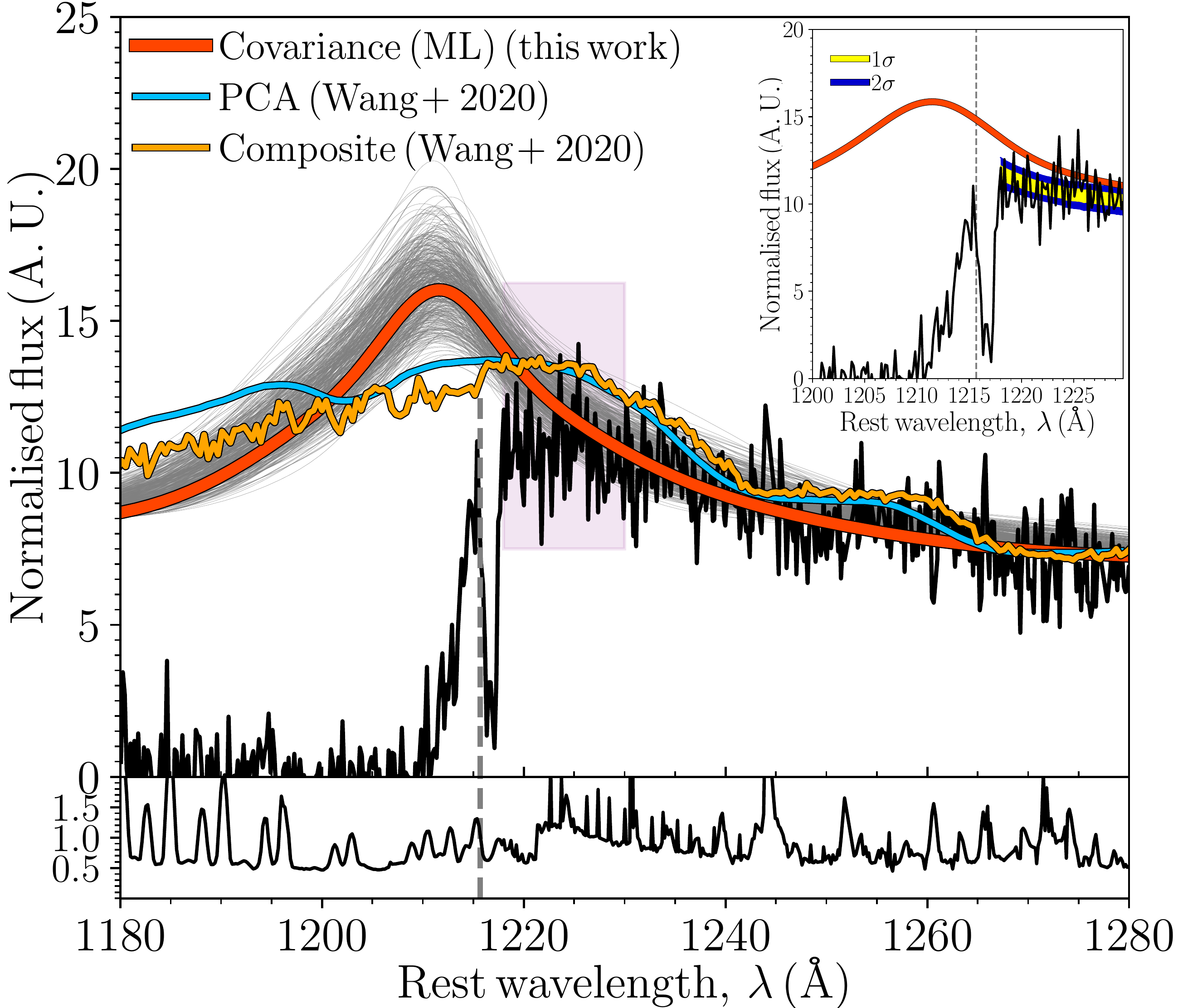}
	\end{center}
\caption[]{The same as Figure~\ref{fig:ReconJ1007}, except for J0252 at $z=7.0$. The black curve (and associated error below) corresponds to the Gemini/GMOS + Keck/NIRES spectrum, the blue line corresponds to the PCA reconstructed profile and the orange line corresponds to a composite spectrum of 83 low-$z$ analogue QSOs \citep{Wang:2020} with large \civ{} blueshifts (3,000 - 5,000 km/s) comparable to J0252.}
\label{fig:ReconJ0252}
\end{figure*}

Next, in Figure~\ref{fig:ReconJ0252}, we provide our reconstructed intrinsic QSO profile for J0252. Again, the best-fit (ML) reconstruction profile is given by the red curve while the thin grey curves highlight a random subsample of 300 reconstruction profiles drawn from the full posterior distribution. In comparison, the blue curve corresponds to the PCA reconstructed profile from \citet{Wang:2020} using the method outlined in \citet{Davies:2018a} while the orange curve is a composite spectrum constructed from 83 low-$z$ (SDSS/BOSS) QSOs selected for their large \civ{} blueshift (3,000 - 5,000 km/s), analogous to J0252. Again, in Figure~\ref{fig:FitJ0252}, we provide the full MCMC fit of J0252 used in our reconstruction pipeline.

Once again, qualitatively speaking, we observe notably different reconstructed profiles from the two approaches. Again, our covariance matrix approach prefers the existence of a strong narrow \lya{} emission line component, peaked around $\lambda\sim1210$\AA. Unlike for J1007, we do not find any preference for any notable \nv{} line component. This differs to the PCA reconstruction in \citet{Wang:2020}, where significant emission is observed between $\lambda\sim1220-1235$\AA. This could be either \lya{} and \nv{} or simply a large amplitude, though strongly blue shifted, \nv{} component. It is harder to ascertain the origins of these features from PCA approaches as they do not predict single peaked emission features. As mentioned earlier, we will discuss differences in the two pipelines in Section~\ref{sec:Discussion}. Importantly though, as was the case for J1007 previously, within the error bounds of both pipelines the recovered profiles appear to be statistically consistent.

Across the region for which we fit the damping wing ($1218$\AA\ $ < \lambda < 1230$\AA), the shape and amplitude of the reconstructed profiles differ considerably. Our reconstructed profiles are systematically lower than those from the PCA reconstruction of \citet{Wang:2020}. In the inset of Figure~\ref{fig:ReconJ0252}, we again show the confidence intervals following our joint fitting procedure (Section~\ref{sec:JointFitting}) with the yellow (blue) shaded regions corresponding to the 68 (95) percentiles. Here, the ML (red) curve notably deviates from these shaded regions indicative clear evidence of notable attenuation of the intrinsic flux by a neutral IGM.

\subsection{Recovered IGM damping wing constraints}

\begin{figure*} 
	\begin{center}
	  \includegraphics[trim = 0.3cm 0.6cm 0cm 0.5cm, scale = 0.54]{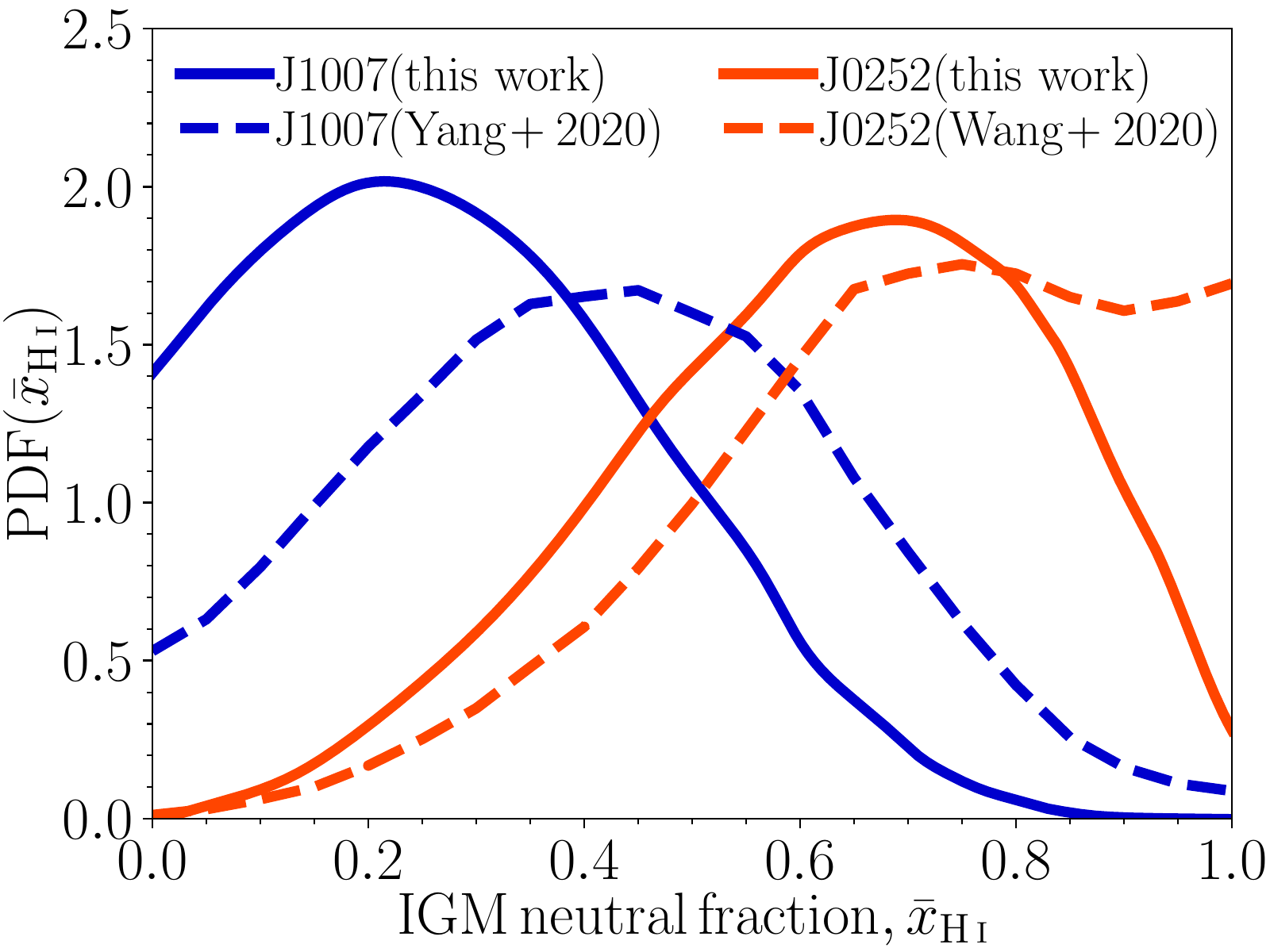}
	  \includegraphics[trim = 0cm 0.6cm 0cm 0.5cm, scale = 0.54]{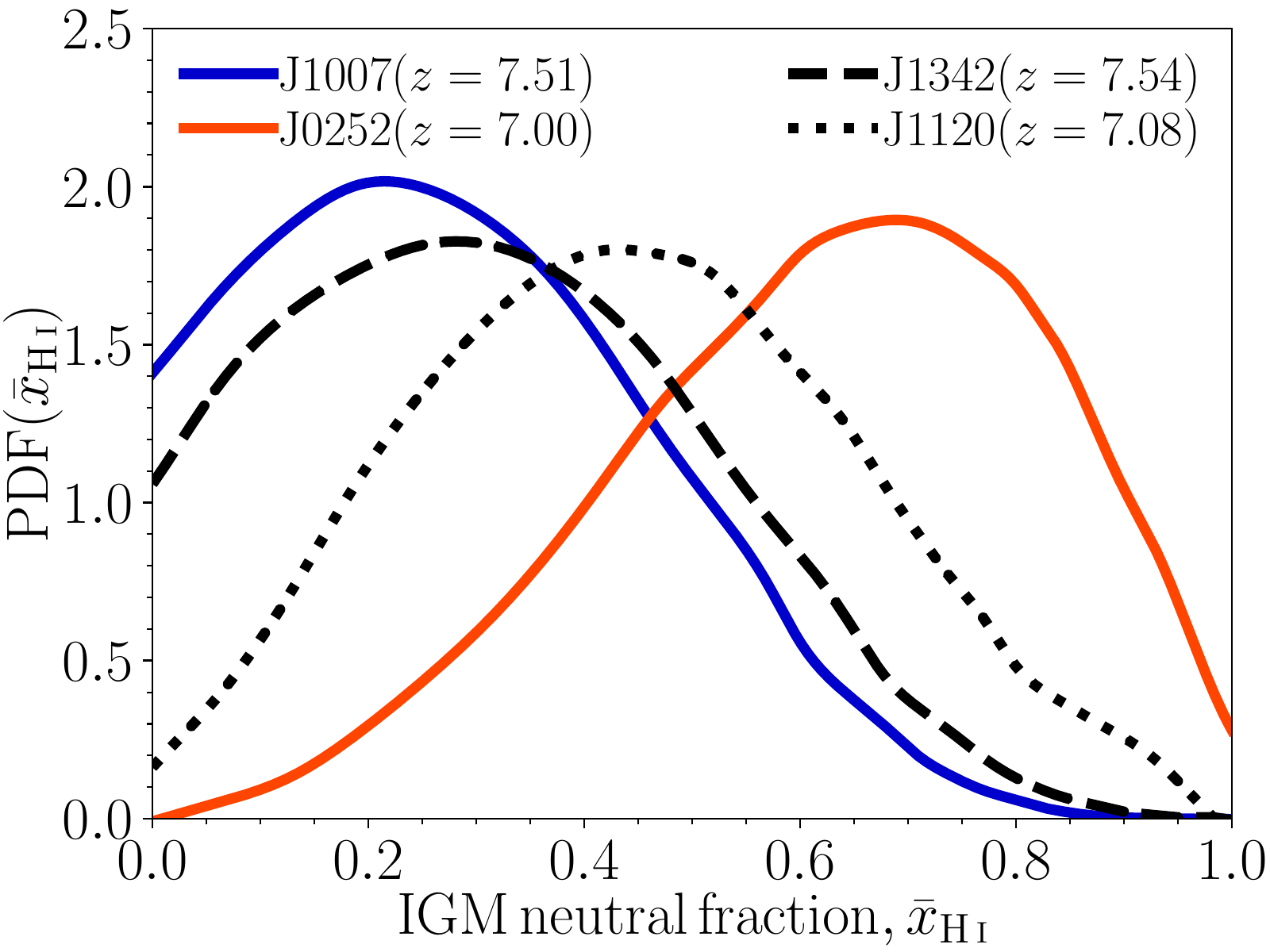}
	\end{center}
\caption[]{The marginalised 1D PDFs of the IGM neutral fraction. \textit{Left panel:} Comparison of the recovered constraints on the IGM neutral fraction for J0252 at $z=7.00$ (red curve) and J1007 at $z=7.51$ (blue curve). Solid curves denote constraints from this work (covariance matrix reconstruction) whereas dashed curves correspond to previous results using PCA reconstruction, \citet{Wang:2020} and \citet{Yang:2020}. \textit{Right panel:} Compilation of the IGM neutral fraction constraints from all $z\gtrsim7$ QSOs using covariance matrix reconstruction (including \nv{}) and the \intermediateHII\ EoR morphology.}
\label{fig:PDFs}
\end{figure*}

In the left panel of Figure~\ref{fig:PDFs} we present the 1D PDFs of our IGM neutral fraction constraints from our damping wing fits to J1007 (blue curve) and J0252 (red curve) using the \intermediateHII\ EoR morphology. For comparison, we additionally show the 1D PDFs recovered by the PCA approach of \citet{Davies:2018a} presented by \citet{Yang:2020} (blue dashed curve) and \citet{Wang:2020} (red dashed curve).

Quantitatively we recover IGM neutral fraction constraints (with 68 per cent confidence limits) of:
\begin{itemize}
\item $\bar{x}_{\hi{}} = 0.27\substack{+0.21 \\ -0.17}$ at $z=7.51$ for J1007
\item $\bar{x}_{\hi{}} = 0.64\substack{+0.19 \\ -0.23}$ at $z=7.00$ for J0252.
\end{itemize}
In comparison, for J1007 \citet{Yang:2020} found $\bar{x}_{\hi{}} = 0.39\substack{+0.22 \\ -0.13}$ and for J0252 \citet{Wang:2020} recovered $\bar{x}_{\hi{}} = 0.70\substack{+0.20 \\ -0.23}$. For both QSOs, the results in this work recover neutral fraction systematically lower than those recovered from the PCA approach. However, these differences are relatively minor, given the relatively large confidence limits. Quantitatively, our results are $\lesssim0.5\sigma$ below the previously reported constraints. In the next section, we will discuss in detail the similarities and differences between the two approaches.

Finally, we also update our constraints on the IGM neutral fraction from the two previously known $z\gtrsim7$ QSOs following the extension of our covariance matrix reconstruction pipeline to include the \nv{} emission line. Specifically, these are ULASJ1120+0641 \citep[hereafter J1120;][]{Mortlock:2011p1049} at $z=7.08$ and ULASJ1342+0928 (hereafter J1342; \citealt{Banados:2018}) at $z=7.54$. Previously, using the same \intermediateHII\ EoR morphology, we recovered $\bar{x}_{\hi{}} = 0.40\substack{+0.21 \\ -0.19}$ for J1120 and $\bar{x}_{\hi{}} = 0.21\substack{+0.17 \\ -0.19}$ for J1342.

In the right panel of Figure~\ref{fig:PDFs} we present a compilation of the IGM neutral fraction constraints for of all four known $z\gtrsim7$ QSOs using our covariance matrix approach (with \nv{}) assuming the \intermediateHII\ EoR morphology. The red and blue curves correspond to J0252 and J1007 as per the right panel of Figure~\ref{fig:PDFs}, whereas the black dotted and dashed curves correspond to the new constraints on the IGM neutral fraction from J1120 and J1342, respectively. Quantitatively, following the inclusion of \nv{} we now update our constraints to the following:
\begin{itemize}
\item $\bar{x}_{\hi{}} = 0.44\substack{+0.23 \\ -0.24}$ at $z=7.08$ for J1120
\item $\bar{x}_{\hi{}} = 0.31\substack{+0.18 \\ -0.19}$ at $z=7.54$ for J1342.
\end{itemize}
For both, we find a higher \igm{} neutral fraction owing to an overall increase in the predicted intrinsic flux following the inclusion of the \nv{} line. Further, we also note an increase to the 68 per cent confidence intervals, owing to the increased scatter in the reconstructed profiles going from a 6 dimensional covariance matrix for the two component of \lya{} (broad and narrow) to the new, 9 dimensional covariance matrix jointly reconstructing \nv{}. In Appendix~\ref{sec:reanalysis} we provide the updated reconstructed profiles for J1120 (Figure~\ref{fig:ULASJ1120}) and J1342 (Figure~\ref{fig:ULASJ1342}) along with a discussion of the new profiles following the inclusion of \nv{} into our analysis pipeline.

\subsection{Compilation of reionisation constraints}

We now have IGM damping wing constraints on four $z\gtrsim7$ QSOs obtained from two distinctly different reconstruction methods and damping wing analyses, as discussed in earlier sections. Here, we combine all of these constraints to obtain one, unified constraint on the IGM neutral fraction from the IGM damping wing imprint\footnote{Note here that we specifically focus on damping wing analyses that consider an inhomogeneous IGM. That is we do not include the constraints on J1120 or J1342 from \citet{Banados:2018,Dominika:2020} or \citet{Reiman:2020}.}. For this, we first average the two individual neutral fraction PDFs (corresponding to the two separate reconstruction pipelines; covariance matrix \citealt{Greig:2017a} and PCA \citealt{Davies:2018a}) for each of the four QSOs. In doing so we in effect average out the inherent systematics from the two different approaches. We then treat these four averaged results as independent measurements on the IGM neutral fraction, allowing us to multiply each individual likelihood to obtain a single, joint posterior for the IGM neutral fraction. Following this procedure, we obtain:
\begin{itemize}
\item $\bar{x}_{\hi{}} = 0.49\substack{+0.13 \\ -0.14}$ at $z=7.29\pm 0.27$.
\end{itemize}
Note, in collapsing these constraints into a single datapoint we are in effect conservatively averaging over all modelling differences and systematics. Further, as these four QSOs span a redshift range of $\Delta z\sim0.5$, cosmic evolution across all these QSOs should be fairly modest.

\begin{figure*} 
	\begin{center}
	  \includegraphics[trim = 0.6cm 0.6cm 0.2cm 0.0cm, scale = 0.75]{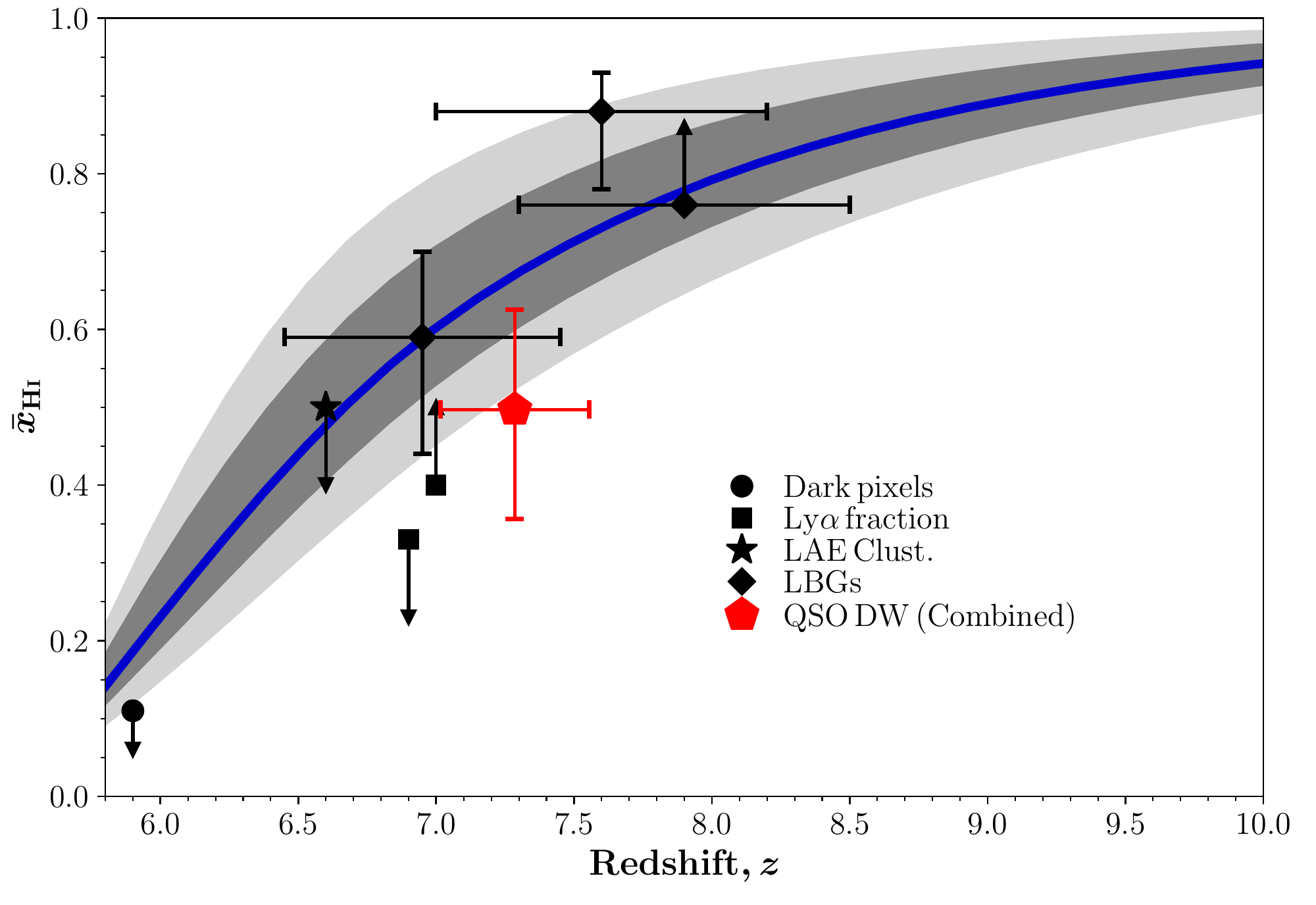}
	\end{center}
\caption[]{A compilation of existing constraints on the IGM neutral fraction as a function of redshift. \textit{Circles:} Dark pixels at $z=5.9$ \citep{McGreer:2015p3668}, \textit{Squares:} the \lya{} fraction at $z=6.9$ \citep{Wold:2021} and $z=7$ \citep{Mesinger:2015p1584}, \textit{Stars:} LAE clustering at $z=6.6$ \citep{Sobacchi:2015}, \textit{Diamonds:} LBGs at $z=7$ \citep{Mason:2018}, $z=7.6$ \citep{Hoag:2019} and $z=8$ \citep{Mason:2019}. The red pentagon corresponds to the combined constraints of all four $z\sim7$ QSOs considered in this work. The blue curve and the dark and light shaded regions corresponds to the median, 1 and 2$\sigma$ constraints from forward-modelling: (i) large-scale \lya{} forest opacity fluctuations; (ii) the CMB optical depth, $\tau_{\rm e}$; (iii) the fraction of dark pixels in the \lya{} and \lyb{} forests and (iv) observed UV LFs at $z=6-10$ \citep{Qin:2021}.}
\label{fig:Hist}
\end{figure*}

In Figure~\ref{fig:Hist} we place this unified QSO damping wing constraint (red pentagon) in context with other constraints on the IGM neutral fraction during reionisation. Here, we consider constraints and limits obtained from: (i) dark pixels \citep{McGreer:2015p3668}, (ii) \lya{} fraction at $z=6.9$ \citep{Wold:2021} and at $z=7$ \citep{Mesinger:2015p1584}, (iii) the clustering of \lya{} emitters (LAEs) at $z=6.6$ \citep{Sobacchi:2015} and (iv) Lyman-break galaxies (LBGs) at $z=7$ \citep{Mason:2018}, $z=7.6$ \citep{Hoag:2019} and at $z=8$ \citep{Mason:2019}. Additionally, we provide constraints on the reionisation history obtained from a Monte-Carlo Markov Chain (MCMC) analysis of the simulated 21-cm signal constrained by existing observations of the reionisation epoch \citep{Qin:2021}. Specifically, these models are constrained by observed UV galaxy LFs at $z=6-10$, the electron scattering optical depth, $\tau_{\rm e}$, measured by Planck \citep{Planck:2018}, the dark pixel limits on the IGM neutral fraction \citep{McGreer:2015p3668} and PDFs of the \lya{} effective optical depth from the \lya{} forest at $z=5-6$ \citep{Bosman:2018}. The median reionisation history is represented by the blue line, whereas the dark and light grey shaded regions correspond to the 1 and 2$\sigma$ confidence intervals. 

This unified QSO datapoint implies a mid-point of reionisation at $z\sim7.3$, slightly below similar limits and constraints from LBGs. However, within the appreciable $1\sigma$ uncertainties they are consistent. With respect to the observationally constrained reionisation histories extracted by forward modelling \lya{} forest data in a Bayesian framework by \citet{Qin:2021}, the median QSO damping wing constraint is $2-3\sigma$ below the median reionisation history, however, again owing to the relatively large uncertainties in averaging across all QSOs, it is still consistent within error. This lower amplitude constraint from the combined QSO damping wing is driven by both $z\sim7.5$ QSOs, which all individually sit below these $2\sigma$ reionisation histories. Interestingly, the posteriors from \citet{Qin:2021} did not consider any of the QSO data points shown in Figure~\ref{fig:Hist} (except for the upper limits from dark pixels), thus it is reassuring that these constraints are consistent.

\section{Discussion} \label{sec:Discussion}

In this work, we have updated the covariance matrix reconstruction pipeline originally outlined in \citet{Greig:2017a} to include a \nv{} emission line component and to extract independent constraints on the IGM neutral fraction using our own methodology \citep{Greig:2017,Greig:2019}. This serves as a complimentary analysis to that performed in \citet{Yang:2020} for J1007 and \citet{Wang:2020} for J0252 using the PCA based reconstruction and damping wing fitting method of \cite{Davies:2018a,Davies:2018}. Although we have recovered consistent constraints on the IGM neutral fraction between the two methods, these have been achieved following various different assumptions for which we outline below.

At their core, the covariance matrix reconstruction pipeline \citep{Greig:2017a} used in this work and the PCA approach from \cite{Davies:2018a} are very similar. Both, are based on similar\footnote{In \citep{Greig:2017a} the training set was limited to QSOs with S/N~$>15$, whereas in \cite{Davies:2018a} it was S/N~$>7$. As a result, after the removal of contaminants the training set sizes were 1653 and 12,764 respectively.} training sets of BOSS QSOs spanning $z\sim2.1-2.5$ and use learnt correlations in the properties of the QSO spectrum redward of \lya{} to reconstruct the intrinsic \lya{} profile. In \citet{Greig:2017a}, it is assumed that all the correlations between the measured Gaussian emission line parameters can be fully described by its covariance matrix (i.e. that it is normally distributed). In \cite{Davies:2018a}, the correlations are instead measured amongst the decomposed principal components of the observed spectra. In utilising principal components, this reduces the dimensionality of the reconstruction pipeline. For example, in this work we have a 21 dimensional covariance matrix that collapses into a 9 dimensional matrix in order to perform the reconstruction, whereas in \cite{Davies:2018a} correlations are linearised between 10 principal components on the red-side and 6 on the blue-side. Add to this the fact that our training set is an order of magnitude smaller to that of \cite{Davies:2018a} and that not all emission line parameters strongly covary amongst one another, we expect the uncertainties (scatter) in reconstructed profiles to be larger in our reconstruction pipeline.

Further, this choice to reconstruct on a series of emission line parameters defined by Gaussian profiles or principal components results in notably different shapes for the predicted intrinsic profiles. Our method, based on the three Gaussian emission line components (two for \lya{} and one for \nv{}) will have a relatively simple and smooth shape. On the other hand, for the PCA approach the shape is not tied to predicting individual components, rather it is the weighted sum of the several predicted blueward principal components. As such the predicted profile following the PCA approach can result in additional features not reproducible by our Gaussian profiles. Thus the overall shape and amplitude of the two pipelines will always differ making direct comparisons less trivial. This difference additionally extends into the distribution of posteriors drawn around the maximum likelihood profile across the two pipelines. While the qualitative trends will be similar across both pipelines, our covariance matrix approach can trivially extract meaningful posterior profiles representing realistic spectra, whereas the PCA approach cannot as its posteriors are generated via perturbations around the maximum likelihood principal components. Thus, the posterior profiles extracted from the PCA approach have less physical meaning than those from the covariance matrix approach. However, while the PCA posteriors may not have physical shapes, one advantage of this approach are that they are drawn from errors from actual spectra rather than from errors in fitted Gaussian profiles.

Equally, the methodology for extracting the IGM damping wing constraints is distinctly different between the two approaches. Our pipeline specifically focuses on avoiding the detailed modelling of the QSO host environment by only fitting for the damping wing imprint sufficiently redward of \lya{} to avoid signatures of inflowing gas (motivating our fitting range of $\lambda=1218-1230$\AA). This approach considerably minimises modelling uncertainties. Whereas in \cite{Davies:2018}, they use a hybrid scheme of numerical simulations to account for uncertainties within the QSO environment to allow them to utilise a much broader fitting region ($\lambda\sim1190-1230$\AA, with the exact range differing on a per QSO basis).

Both approaches use large-scale semi-numerical simulations (i.e. \cmfst{} \citealt{Mesinger:2011p1123}; though with different physical prescriptions) to extract the smooth damping wing profiles from interactions with neutral hydrogen along the line-of-sight. However, by additionally modelling blueward of \lya{} \cite{Davies:2018} must also use a hybrid scheme of hydrodynamical simulations to model the density environment of the QSO host coupled with 1D radiative transfer simulations to model the impact of the QSO flux on its nearby environment. As such, this adds in the additional complexity of the QSO ionising contribution to the environment, which can be somewhat degenerate with the IGM neutral fraction and must be marginalised over when obtaining constraints from the IGM damping wing imprint. Therefore, while this approach has the advantage of fitting for the damping wing imprint over a broader spatial range, there are many additional complexities and uncertainties embedded in the models. On the other hand, by restricting ourselves to only considering redward of \lya{} our constraints can be considered more model independent. However, our approach does assume a prior on the minimum \hii{} region size surrounding the QSO, which effectively translates into a non-trivial prior on the QSO's ionising contribution to it's environment. Such an ad-hoc prior could bias results towards larger QSO lifetimes when the neutral fraction approaches unity and when the likelihood is not strongly-constraining (see \citealt{Greig:2019} for further details).

Importantly, despite the notable differences in the modelling and methodology of the two pipelines, in this work we have found similar constraints on the \igm{} neutral fraction from both J1007 and J0252. However, this is not always the case. For example, for J1342, even despite the inclusion of \nv{} increasing our recovered constraints on the IGM neutral fraction from this QSO, we still differ by $\sim1\sigma$. In order to get a true handle on the similarities, differences and any potential biases inherent in either pipeline when applied to these high-$z$ QSOs, it will be useful to perform a detailed comparison on a unified set of QSOs. However, we leave such an exploration to future work.

\section{Conclusion} \label{sec:Conclusion}

In this work, we perform an independent and complimentary IGM damping wing analysis on two recently discovered $z\gtrsim7$~QSOs; J1007 \citep{Yang:2020} at $z=7.51$ and J0252 \citep{Wang:2020} at $z=7.00$. Additionally, we update the existing Gaussian covariance matrix approach to predict the intrinsic QSO profile developed by \citet{Greig:2017} to additionally reconstruct the \nv{} emission line. Using this modified reconstruction pipeline we then extract the damping wing imprint of the neutral IGM by jointly fitting the predicted intrinsic QSO flux with synthetic damping wing profiles drawn from realistic EoR simulations to the observed spectra of the $z\gtrsim7$QSOs.

Specifically, we fit the emission lines of the observed QSO spectrum over $\lambda=1275-2300$\AA\ and evaluate our measured covariance matrix of various known emission lines. In doing so, we obtain a reduced covariance matrix (containing \lya{} and \nv{} emission line parameters) from which we extract $10^5$ reconstructed intrinsic QSO profiles. We then combine these intrinsic profiles with $10^5$ synthetic damping wing profiles drawn from large-volume EoR simulation (1.6 Gpc on a side with an EoR morphology driven by galaxies residing in $M_h \gsim 10^9 M_\odot$ haloes, in our nomenclature referred to as an \intermediateHII\ model). We then jointly fit these $10^{10}$ profiles to the observed spectrum of J1007 and J0252 redward of \lya{} ($\lambda=1218-1230$\AA) in a Bayesian framework. Following this procedure, we recover the following constraints on the IGM neutral fraction (with 68 per cent confidence intervals):
\begin{itemize}
\item $\bar{x}_{\hi{}} = 0.27\substack{+0.21 \\ -0.17}$ at $z=7.51$ for J1007
\item $\bar{x}_{\hi{}} = 0.64\substack{+0.19 \\ -0.23}$ at $z=7.00$ for J0252.
\end{itemize}
These constraints are consistent (within error), albeit slightly lower than those previously reported in the literature; $\bar{x}_{\hi{}} = 0.39\substack{+0.22 \\ -0.13}$ for J1007 \citep{Yang:2020} and $\bar{x}_{\hi{}} = 0.70\substack{+0.20 \\ -0.23}$ for J0252 \citep{Wang:2020}. These can be attributed to the subtleties of the two methodologies: (i) our covariance matrix approach reconstructs multiple Gaussian emission profiles whereas the other works reconstruct on principal components (PCA) resulting in different profile shapes and (ii) we fit the damping wing only redward of \lya{} ($\lambda=1218-1230$\AA) whereas the other works additionally fit blueward of \lya{} requiring additional modelling of the QSO host environment.

Additionally, since we have updated the covariance matrix reconstruction approach to include the \nv{} emission line, we perform a reanalysis of the IGM damping wing constraints from both J1120 \citep{Mortlock:2011p1049} at $z=7.08$ and J1342 \citep{Banados:2018} at $z=7.54$. After including \nv{}, under the \intermediateHII\ EoR morphology we find:
\begin{itemize}
\item $\bar{x}_{\hi{}} = 0.44\substack{+0.23 \\ -0.24}$ at $z=7.08$ for J1120
\item $\bar{x}_{\hi{}} = 0.31\substack{+0.18 \\ -0.19}$ at $z=7.54$ for J1342.
\end{itemize}
These are consistent within error to our previously reported results of $\bar{x}_{\hi{}} = 0.40\substack{+0.21 \\ -0.19}$ for J1120 \citep{Greig:2017} and $\bar{x}_{\hi{}} = 0.21\substack{+0.17 \\ -0.19}$ for J1342 \citep{Greig:2019}. Note that for J1120, the differences are minor as in that work we applied a prior on the reconstructed flux down to $\lambda=1230$\AA\ allowing for the fitting of \nv{}. For J1342, the increase in the IGM neutral fraction arises from the \nv{} line component, and slightly diminishes the tension with the reported results from \citet{Davies:2018} from $\sim1.5\sigma$ to $\sim1\sigma$.

Finally, we combined the constraints on the IGM neutral fraction from all four known $z\gtrsim7$ QSOs across the two analysis pipelines (covariance matrix, used in this work and the PCA approach in the literature) to obtain:
\begin{itemize}
\item $\bar{x}_{\hi{}} = 0.49\substack{+0.11 \\ -0.11}$ at $z=7.29$.
\end{itemize}
We found this IGM neutral fraction, with the relatively broad uncertainties obtained from averaging across all four $z\gtrsim7$ QSOs, to be roughly consistent to within $\sim1\sigma$ of the median constraints on the IGM neutral fraction obtained by forward-modelling recent \lya{} forest data by \citet{Qin:2021}.

\section*{Acknowledgements}

We thank Aaron Barth for providing the Keck NIRES data and both Aaron Barth and Xiaohiu Fan for reading an early version of this manuscript. We thank Yuxiang Qin for providing the reionisation history posteriors. Parts of this research were supported by the Australian Research Council Centre of Excellence for All Sky Astrophysics in 3 Dimensions (ASTRO 3D), through project number CE170100013. A. M. acknowledges funding from the European Research Council (ERC) under the European Union's Horizon 2020 research and innovation programme (grant agreement No 638809 -- AIDA). The results presented here reflect the authors' views; the ERC is not responsible for their use. FW acknowledge the support provided by NASA through the NASA Hubble Fellowship grant \#HST-HF2- 51448.001-A awarded by the Space Telescope Science Institute, which is operated by the Association of Universities for Research in Astronomy, Incorporated, under NASA contract NAS5-26555.

\section*{Data Availability}

The data underlying this article will be shared on reasonable request to the corresponding author.

\bibliography{Papers}

\begin{thebibliography}{}
\makeatletter
\relax
\def\mn@urlcharsother{\let\do\@makeother \do\$\do\&\do\#\do\^\do\_\do\%\do\~}
\def\mn@doi{\begingroup\mn@urlcharsother \@ifnextchar [ {\mn@doi@}
  {\mn@doi@[]}}
\def\mn@doi@[#1]#2{\def\@tempa{#1}\ifx\@tempa\@empty \href
  {http://dx.doi.org/#2} {doi:#2}\else \href {http://dx.doi.org/#2} {#1}\fi
  \endgroup}
\def\mn@eprint#1#2{\mn@eprint@#1:#2::\@nil}
\def\mn@eprint@arXiv#1{\href {http://arxiv.org/abs/#1} {{\tt arXiv:#1}}}
\def\mn@eprint@dblp#1{\href {http://dblp.uni-trier.de/rec/bibtex/#1.xml}
  {dblp:#1}}
\def\mn@eprint@#1:#2:#3:#4\@nil{\def\@tempa {#1}\def\@tempb {#2}\def\@tempc
  {#3}\ifx \@tempc \@empty \let \@tempc \@tempb \let \@tempb \@tempa \fi \ifx
  \@tempb \@empty \def\@tempb {arXiv}\fi \@ifundefined
  {mn@eprint@\@tempb}{\@tempb:\@tempc}{\expandafter \expandafter \csname
  mn@eprint@\@tempb\endcsname \expandafter{\@tempc}}}

\bibitem[\protect\citeauthoryear{Alam et~al.}{Alam
  et~al.}{2015}]{Alam:2015p5162}
Alam S.,  et~al., 2015, \mn@doi [ApJS] {10.1088/0067-0049/219/1/12}, \href
  {https://ui.adsabs.harvard.edu/abs/2015ApJS..219...12A} {219, 12}

\bibitem[\protect\citeauthoryear{Ba{\~n}ados et~al.}{Ba{\~n}ados
  et~al.}{2018}]{Banados:2018}
Ba{\~n}ados E.,  et~al., 2018, \mn@doi [Nature] {10.1038/nature25180}, \href
  {https://ui.adsabs.harvard.edu/abs/2018Natur.553..473B} {553, 473}

\bibitem[\protect\citeauthoryear{{Becker}, {Bolton}, {Madau}, {Pettini},
  {Ryan-Weber}  \& {Venemans}}{{Becker} et~al.}{2015}]{Becker:2015}
{Becker} G.~D.,  {Bolton} J.~S.,  {Madau} P.,  {Pettini} M.,  {Ryan-Weber}
  E.~V.,   {Venemans} B.~P.,  2015, \mn@doi [MNRAS] {10.1093/mnras/stu2646},
  \href {https://ui.adsabs.harvard.edu/abs/2015MNRAS.447.3402B} {447, 3402}

\bibitem[\protect\citeauthoryear{Bolton, Haehnelt, Warren, Hewett, Mortlock,
  Venemans, McMahon  \& Simpson}{Bolton et~al.}{2011}]{Bolton:2011p1063}
Bolton J.~S.,  Haehnelt M.~G.,  Warren S.~J.,  Hewett P.~C.,  Mortlock D.~J.,
  Venemans B.~P.,  McMahon R.~G.,   Simpson C.,  2011, \mn@doi [MNRAS]
  {10.1111/j.1745-3933.2011.01100.x}, \href
  {https://ui.adsabs.harvard.edu/abs/2011MNRAS.416L..70B} {416, L70}

\bibitem[\protect\citeauthoryear{{Bosman} \& {Becker}}{{Bosman} \&
  {Becker}}{2015}]{Bosman:2015}
{Bosman} S. E.~I.,  {Becker} G.~D.,  2015, \mn@doi [\mnras]
  {10.1093/mnras/stv1336}, \href
  {https://ui.adsabs.harvard.edu/abs/2015MNRAS.452.1105B} {452, 1105}

\bibitem[\protect\citeauthoryear{{Bosman}, {Fan}, {Jiang}, {Reed}, {Matsuoka},
  {Becker}  \& {Haehnelt}}{{Bosman} et~al.}{2018}]{Bosman:2018}
{Bosman} S. E.~I.,  {Fan} X.,  {Jiang} L.,  {Reed} S.,  {Matsuoka} Y.,
  {Becker} G.,   {Haehnelt} M.,  2018, \mn@doi [MNRAS] {10.1093/mnras/sty1344},
  \href {https://ui.adsabs.harvard.edu/abs/2018MNRAS.479.1055B} {479, 1055}

\bibitem[\protect\citeauthoryear{{Bosman} et~al.,}{{Bosman}
  et~al.}{2021a}]{Bosman:2021a}
{Bosman} S. E.~I.,  et~al., 2021a, arXiv e-prints, \href
  {https://ui.adsabs.harvard.edu/abs/2021arXiv210803699B} {p. arXiv:2108.03699}

\bibitem[\protect\citeauthoryear{{Bosman}, {{\v{D}}urov{\v{c}}{\'\i}kov{\'a}},
  {Davies}  \& {Eilers}}{{Bosman} et~al.}{2021b}]{Bosman:2021}
{Bosman} S. E.~I.,  {{\v{D}}urov{\v{c}}{\'\i}kov{\'a}} D.,  {Davies} F.~B.,
  {Eilers} A.-C.,  2021b, \mn@doi [\mnras] {10.1093/mnras/stab572}, \href
  {https://ui.adsabs.harvard.edu/abs/2021MNRAS.503.2077B} {503, 2077}

\bibitem[\protect\citeauthoryear{{Choudhury}, {Paranjape}  \&
  {Bosman}}{{Choudhury} et~al.}{2021}]{Choudhury:2021}
{Choudhury} T.~R.,  {Paranjape} A.,   {Bosman} S. E.~I.,  2021, \mn@doi [MNRAS]
  {10.1093/mnras/stab045}, \href
  {https://ui.adsabs.harvard.edu/abs/2021MNRAS.501.5782C} {501, 5782}

\bibitem[\protect\citeauthoryear{{Davies} et~al.}{{Davies}
  et~al.}{2018a}]{Davies:2018}
{Davies} F.~B.,  et~al., 2018a, \mn@doi [ApJ] {10.3847/1538-4357/aad6dc}, \href
  {https://ui.adsabs.harvard.edu/abs/2018ApJ...864..142D} {864, 142}

\bibitem[\protect\citeauthoryear{{Davies} et~al.,}{{Davies}
  et~al.}{2018b}]{Davies:2018a}
{Davies} F.~B.,  et~al., 2018b, \mn@doi [ApJ] {10.3847/1538-4357/aad7f8}, \href
  {https://ui.adsabs.harvard.edu/abs/2018ApJ...864..143D} {864, 143}

\bibitem[\protect\citeauthoryear{Dawson et~al.}{Dawson
  et~al.}{2013}]{Dawson:2013p5160}
Dawson K.~S.,  et~al., 2013, \mn@doi [AJ] {10.1088/0004-6256/145/1/10}, \href
  {https://ui.adsabs.harvard.edu/abs/2013AJ....145...10D} {145, 10}

\bibitem[\protect\citeauthoryear{{Eilers}, {Davies}  \& {Hennawi}}{{Eilers}
  et~al.}{2018}]{Eilers:2018}
{Eilers} A.-C.,  {Davies} F.~B.,   {Hennawi} J.~F.,  2018, \mn@doi [\apj]
  {10.3847/1538-4357/aad4fd}, \href
  {https://ui.adsabs.harvard.edu/abs/2018ApJ...864...53E} {864, 53}

\bibitem[\protect\citeauthoryear{Fan et~al.}{Fan et~al.}{2006}]{Fan:2006p4005}
Fan X.,  et~al., 2006, \mn@doi [AJ] {10.1086/504836}, \href
  {https://ui.adsabs.harvard.edu/abs/2006AJ....132..117F} {132, 117}

\bibitem[\protect\citeauthoryear{{Fathivavsari}}{{Fathivavsari}}{2020}]{Fathivavsari:2020}
{Fathivavsari} H.,  2020, \mn@doi [\apj] {10.3847/1538-4357/ab9b7d}, \href
  {https://ui.adsabs.harvard.edu/abs/2020ApJ...898..114F} {898, 114}

\bibitem[\protect\citeauthoryear{{Greig}, {Mesinger}, {McGreer}, {Gallerani}
  \& {Haiman}}{{Greig} et~al.}{2017a}]{Greig:2017a}
{Greig} B.,  {Mesinger} A.,  {McGreer} I.~D.,  {Gallerani} S.,   {Haiman} Z.,
  2017a, \mn@doi [MNRAS] {10.1093/mnras/stw3210}, \href
  {https://ui.adsabs.harvard.edu/abs/2017MNRAS.466.1814G} {466, 1814}

\bibitem[\protect\citeauthoryear{{Greig}, {Mesinger}, {Haiman}  \&
  {Simcoe}}{{Greig} et~al.}{2017b}]{Greig:2017}
{Greig} B.,  {Mesinger} A.,  {Haiman} Z.,   {Simcoe} R.~A.,  2017b, \mn@doi
  [MNRAS] {10.1093/mnras/stw3351}, \href
  {https://ui.adsabs.harvard.edu/abs/2017MNRAS.466.4239G} {466, 4239}

\bibitem[\protect\citeauthoryear{{Greig}, {Mesinger}  \& {Ba{\~n}ados}}{{Greig}
  et~al.}{2019}]{Greig:2019}
{Greig} B.,  {Mesinger} A.,   {Ba{\~n}ados} E.,  2019, \mn@doi [MNRAS]
  {10.1093/mnras/stz230}, \href
  {https://ui.adsabs.harvard.edu/abs/2019MNRAS.484.5094G} {484, 5094}

\bibitem[\protect\citeauthoryear{{Hoag} et~al.}{{Hoag}
  et~al.}{2019}]{Hoag:2019}
{Hoag} A.,  et~al., 2019, \mn@doi [ApJ] {10.3847/1538-4357/ab1de7}, \href
  {https://ui.adsabs.harvard.edu/abs/2019ApJ...878...12H} {878, 12}

\bibitem[\protect\citeauthoryear{{Keating}, {Weinberger}, {Kulkarni},
  {Haehnelt}, {Chardin}  \& {Aubert}}{{Keating} et~al.}{2020a}]{Keating:2020a}
{Keating} L.~C.,  {Weinberger} L.~H.,  {Kulkarni} G.,  {Haehnelt} M.~G.,
  {Chardin} J.,   {Aubert} D.,  2020a, \mn@doi [MNRAS] {10.1093/mnras/stz3083},
  \href {https://ui.adsabs.harvard.edu/abs/2020MNRAS.491.1736K} {491, 1736}

\bibitem[\protect\citeauthoryear{{Keating}, {Kulkarni}, {Haehnelt}, {Chardin}
  \& {Aubert}}{{Keating} et~al.}{2020b}]{Keating:2020b}
{Keating} L.~C.,  {Kulkarni} G.,  {Haehnelt} M.~G.,  {Chardin} J.,   {Aubert}
  D.,  2020b, \mn@doi [MNRAS] {10.1093/mnras/staa1909}, \href
  {https://ui.adsabs.harvard.edu/abs/2020MNRAS.497..906K} {497, 906}

\bibitem[\protect\citeauthoryear{{Kulkarni}, {Keating}, {Haehnelt}, {Bosman},
  {Puchwein}, {Chardin}  \& {Aubert}}{{Kulkarni} et~al.}{2019}]{Kulkarni:2019}
{Kulkarni} G.,  {Keating} L.~C.,  {Haehnelt} M.~G.,  {Bosman} S. E.~I.,
  {Puchwein} E.,  {Chardin} J.,   {Aubert} D.,  2019, \mn@doi [MNRAS]
  {10.1093/mnrasl/slz025}, \href
  {https://ui.adsabs.harvard.edu/abs/2019MNRAS.485L..24K} {485, L24}

\bibitem[\protect\citeauthoryear{{Liu} \& {Bordoloi}}{{Liu} \&
  {Bordoloi}}{2021}]{Liu:2021}
{Liu} B.,  {Bordoloi} R.,  2021, \mn@doi [\mnras] {10.1093/mnras/stab177},
  \href {https://ui.adsabs.harvard.edu/abs/2021MNRAS.502.3510L} {502, 3510}

\bibitem[\protect\citeauthoryear{{Mason}, {Treu}, {Dijkstra}, {Mesinger},
  {Trenti}, {Pentericci}, {de Barros}  \& {Vanzella}}{{Mason}
  et~al.}{2018}]{Mason:2018}
{Mason} C.~A.,  {Treu} T.,  {Dijkstra} M.,  {Mesinger} A.,  {Trenti} M.,
  {Pentericci} L.,  {de Barros} S.,   {Vanzella} E.,  2018, \mn@doi [ApJ]
  {10.3847/1538-4357/aab0a7}, \href
  {https://ui.adsabs.harvard.edu/abs/2018ApJ...856....2M} {856, 2}

\bibitem[\protect\citeauthoryear{{Mason} et~al.}{{Mason}
  et~al.}{2019}]{Mason:2019}
{Mason} C.~A.,  et~al., 2019, \mn@doi [MNRAS] {10.1093/mnras/stz632}, \href
  {https://ui.adsabs.harvard.edu/abs/2019MNRAS.485.3947M} {485, 3947}

\bibitem[\protect\citeauthoryear{McGreer, Mesinger  \& D'Odorico}{McGreer
  et~al.}{2015}]{McGreer:2015p3668}
McGreer I.~D.,  Mesinger A.,   D'Odorico V.,  2015, \mn@doi [MNRAS]
  {10.1093/mnras/stu2449}, \href
  {https://ui.adsabs.harvard.edu/abs/2015MNRAS.447..499M} {447, 499}

\bibitem[\protect\citeauthoryear{{Mesinger} \& {Haiman}}{{Mesinger} \&
  {Haiman}}{2007}]{Mesinger:2007}
{Mesinger} A.,  {Haiman} Z.,  2007, \mn@doi [ApJ] {10.1086/513688}, \href
  {https://ui.adsabs.harvard.edu/abs/2007ApJ...660..923M} {660, 923}

\bibitem[\protect\citeauthoryear{Mesinger, Furlanetto  \& Cen}{Mesinger
  et~al.}{2011}]{Mesinger:2011p1123}
Mesinger A.,  Furlanetto S.,   Cen R.,  2011, \mn@doi [MNRAS]
  {10.1111/j.1365-2966.2010.17731.x}, \href
  {https://ui.adsabs.harvard.edu/abs/2011MNRAS.411..955M} {411, 955}

\bibitem[\protect\citeauthoryear{Mesinger, Aykutalp, Vanzella, Pentericci,
  Ferrara  \& Dijkstra}{Mesinger et~al.}{2015}]{Mesinger:2015p1584}
Mesinger A.,  Aykutalp A.,  Vanzella E.,  Pentericci L.,  Ferrara A.,
  Dijkstra M.,  2015, \mn@doi [MNRAS] {10.1093/mnras/stu2089}, \href
  {https://ui.adsabs.harvard.edu/abs/2015MNRAS.446..566M} {446, 566}

\bibitem[\protect\citeauthoryear{Mesinger, Greig  \& Sobacchi}{Mesinger
  et~al.}{2016}]{Mesinger:2016p6167}
Mesinger A.,  Greig B.,   Sobacchi E.,  2016, \mn@doi [MNRAS]
  {10.1093/mnras/stw831}, \href
  {https://ui.adsabs.harvard.edu/abs/2016MNRAS.459.2342M} {459, 2342}

\bibitem[\protect\citeauthoryear{Miralda-Escud{\'e}}{Miralda-Escud{\'e}}{1998}]{MiraldaEscude:1998p1041}
Miralda-Escud{\'e} J.,  1998, \mn@doi [ApJ] {10.1086/305799}, \href
  {https://ui.adsabs.harvard.edu/abs/1998ApJ...501...15M} {501, 15}

\bibitem[\protect\citeauthoryear{Mortlock et~al.}{Mortlock
  et~al.}{2011}]{Mortlock:2011p1049}
Mortlock D.~J.,  et~al., 2011, \mn@doi [Nature] {10.1038/nature10159}, \href
  {https://ui.adsabs.harvard.edu/abs/2011Natur.474..616M} {474, 616}

\bibitem[\protect\citeauthoryear{{Nasir} \& {D'Aloisio}}{{Nasir} \&
  {D'Aloisio}}{2020}]{Nasir:2020}
{Nasir} F.,  {D'Aloisio} A.,  2020, \mn@doi [MNRAS] {10.1093/mnras/staa894},
  \href {https://ui.adsabs.harvard.edu/abs/2020MNRAS.494.3080N} {494, 3080}

\bibitem[\protect\citeauthoryear{{Planck Collaboration XIII}}{{Planck
  Collaboration XIII}}{2016}]{PlanckCollaboration:2016p7780}
{Planck Collaboration XIII} 2016, \mn@doi [A\&A] {10.1051/0004-6361/201525830},
  \href {https://ui.adsabs.harvard.edu/abs/2016A&A...594A..13P} {594, A13}

\bibitem[\protect\citeauthoryear{{Planck Collaboration} et~al.}{{Planck
  Collaboration} et~al.}{2020}]{Planck:2018}
{Planck Collaboration} et~al., 2020, \mn@doi [A\&A]
  {10.1051/0004-6361/201833910}, 641, A6

\bibitem[\protect\citeauthoryear{{Qin}, {Mesinger}, {Bosman}  \& {Viel}}{{Qin}
  et~al.}{2021}]{Qin:2021}
{Qin} Y.,  {Mesinger} A.,  {Bosman} S. E.~I.,   {Viel} M.,  2021, arXiv
  e-prints, \href {https://ui.adsabs.harvard.edu/abs/2021arXiv210109033Q} {p.
  arXiv:2101.09033}

\bibitem[\protect\citeauthoryear{{Reiman}, {Tamanas}, {Prochaska}  \&
  {{\v{D}}urov{\v{c}}{\'\i}kov{\'a}}}{{Reiman} et~al.}{2020}]{Reiman:2020}
{Reiman} D.~M.,  {Tamanas} J.,  {Prochaska} J.~X.,
  {{\v{D}}urov{\v{c}}{\'\i}kov{\'a}} D.,  2020, arXiv e-prints, \href
  {https://ui.adsabs.harvard.edu/abs/2020arXiv200600615R} {p. arXiv:2006.00615}

\bibitem[\protect\citeauthoryear{{Rybicki} \& {Lightman}}{{Rybicki} \&
  {Lightman}}{1979}]{Rybicki1979}
{Rybicki} G.~B.,  {Lightman} A.~P.,  1979, {Radiative Processes in
  Astrophysics, Wiley-Interscience, New York.}

\bibitem[\protect\citeauthoryear{Simcoe, Sullivan, Cooksey, Kao, Matejek  \&
  Burgasser}{Simcoe et~al.}{2012}]{Simcoe:2012}
Simcoe R.~A.,  Sullivan P.~W.,  Cooksey K.~L.,  Kao M.~M.,  Matejek M.~S.,
  Burgasser A.~J.,  2012, \mn@doi [Nature] {10.1038/nature11612}, \href
  {https://ui.adsabs.harvard.edu/abs/2012Natur.492...79S} {492, 79}

\bibitem[\protect\citeauthoryear{{Sobacchi} \& {Mesinger}}{{Sobacchi} \&
  {Mesinger}}{2015}]{Sobacchi:2015}
{Sobacchi} E.,  {Mesinger} A.,  2015, \mn@doi [MNRAS] {10.1093/mnras/stv1751},
  \href {https://ui.adsabs.harvard.edu/abs/2015MNRAS.453.1843S} {453, 1843}

\bibitem[\protect\citeauthoryear{{Wang} et~al.,}{{Wang}
  et~al.}{2020}]{Wang:2020}
{Wang} F.,  et~al., 2020, \mn@doi [ApJ] {10.3847/1538-4357/ab8c45}, \href
  {https://ui.adsabs.harvard.edu/abs/2020ApJ...896...23W} {896, 23}

\bibitem[\protect\citeauthoryear{{Wold} et~al.}{{Wold}
  et~al.}{2021}]{Wold:2021}
{Wold} I. G.~B.,  et~al., 2021, arXiv e-prints, \href
  {https://ui.adsabs.harvard.edu/abs/2021arXiv210512191W} {p. arXiv:2105.12191}

\bibitem[\protect\citeauthoryear{{Yang} et~al.,}{{Yang}
  et~al.}{2019}]{Yang:2019}
{Yang} J.,  et~al., 2019, \mn@doi [\aj] {10.3847/1538-3881/ab1be1}, \href
  {https://ui.adsabs.harvard.edu/abs/2019AJ....157..236Y} {157, 236}

\bibitem[\protect\citeauthoryear{{Yang} et~al.,}{{Yang}
  et~al.}{2020a}]{Yang:2020}
{Yang} J.,  et~al., 2020a, \mn@doi [ApJL] {10.3847/2041-8213/ab9c26}, \href
  {https://ui.adsabs.harvard.edu/abs/2020ApJ...897L..14Y} {897, L14}

\bibitem[\protect\citeauthoryear{{Yang} et~al.,}{{Yang}
  et~al.}{2020b}]{Yang:2020a}
{Yang} J.,  et~al., 2020b, \mn@doi [\apj] {10.3847/1538-4357/abbc1b}, \href
  {https://ui.adsabs.harvard.edu/abs/2020ApJ...904...26Y} {904, 26}

\bibitem[\protect\citeauthoryear{{{\v{D}}urov{\v{c}}{\'\i}kov{\'a}}, {Katz},
  {Bosman}, {Davies}, {Devriendt}  \&
  {Slyz}}{{{\v{D}}urov{\v{c}}{\'\i}kov{\'a}} et~al.}{2020}]{Dominika:2020}
{{\v{D}}urov{\v{c}}{\'\i}kov{\'a}} D.,  {Katz} H.,  {Bosman} S. E.~I.,
  {Davies} F.~B.,  {Devriendt} J.,   {Slyz} A.,  2020, \mn@doi [\mnras]
  {10.1093/mnras/staa505}, \href
  {https://ui.adsabs.harvard.edu/abs/2020MNRAS.493.4256D} {493, 4256}

\makeatother
\end{thebibliography}

\appendix

\section[]{Covariance Matrix with \nv{} component} \label{sec:Covariance_wNV}

\begin{figure*} 
	\begin{center}
	  \includegraphics[trim = 0.2cm 0.6cm 0cm 0.5cm, scale = 0.84]{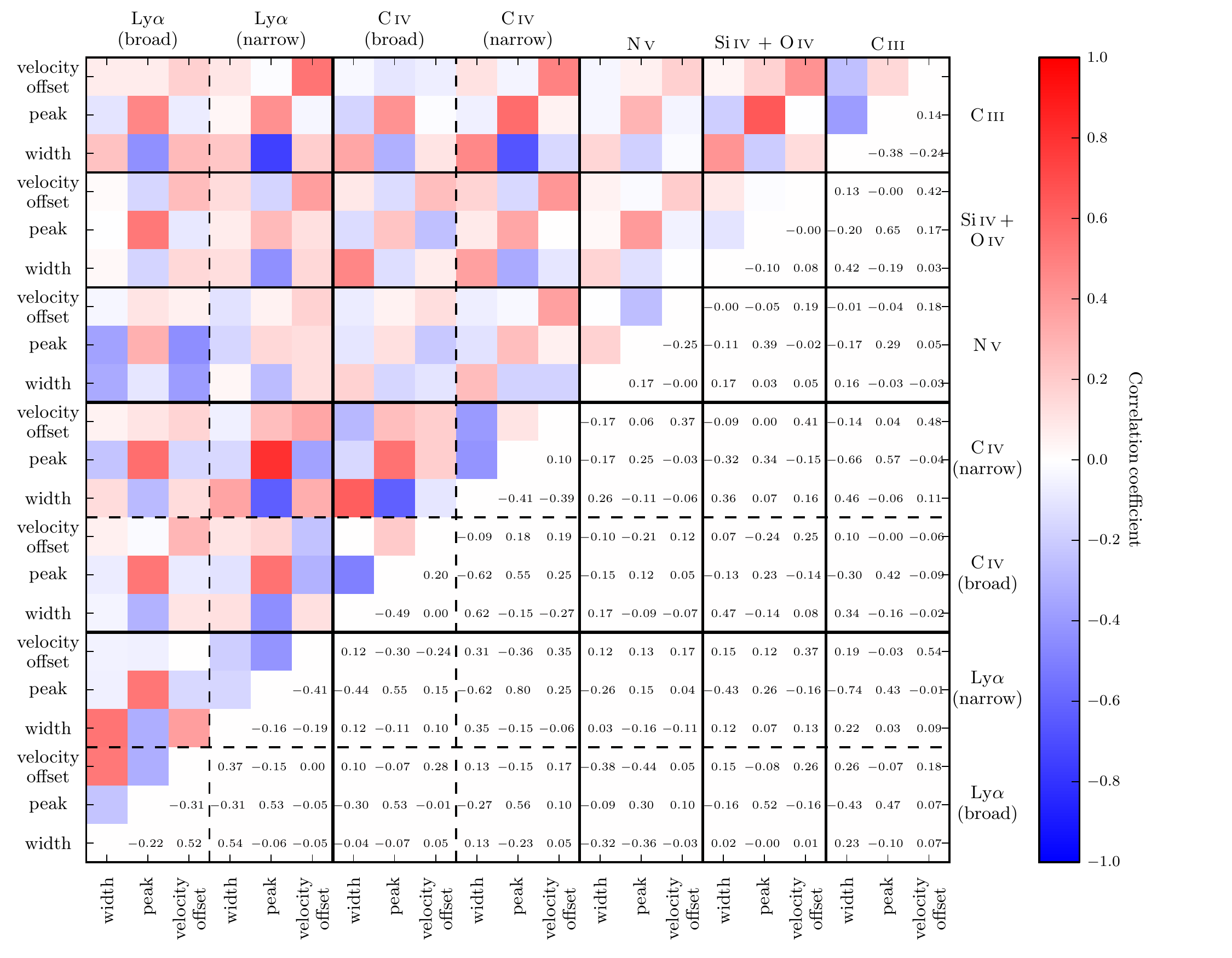}
	\end{center}
\caption[]{The updated correlation coefficient matrix (correlation coefficients listed in the lower half) including the \nv{} component constructed from the `good` sample of 1673 QSOs from \citet{Greig:2017a}. This 21 dimensional matrix contains  double component Gaussians for both \lya{} and \civ{} and single component Gaussians for \nv{}, \sioiv{} and \ciii{}. For each Gaussian line component we have three parameters, the peak width, peak height and velocity offset from systemic.}
\label{fig:CovarianceMatrix}
\end{figure*}

In this work, we have added the \nv{} emission line to the original covariance matrix reconstruction pipeline from \citet{Greig:2017a}. Previously, when performing the Monte Carlo Markov Chain (MCMC) fits to our full QSO training set, we fit for the \nv{} emission line. However, we decided to ignore \nv{} from the covariance matrix as the \nv{} line profile parameters (peak, width and velocity offset) did not strongly correlate with \lya{} or the other primary high ionisation lines (\civ{}, \sioiv{} and \ciii{}) used for our \lya{} reconstruction pipeline. Therefore, to include \nv{} into our analysis we simply updated our covariance matrix using the existing data at hand.

To explore the strength of correlations between the \nv{} line parameters and the other lines, in Figure~\ref{fig:CovarianceMatrix} we present the updated correlation matrix of the 1673 QSOs in the training sample. This correlation coefficient matrix, $\textbfss{R}_{ij}$, is defined as,
\begin{eqnarray}
\textbfss{R}_{ij} = \frac{\textbfss{C}_{ij}}{\sqrt{\textbfss{C}_{ii}\textbfss{C}_{jj}}},
\end{eqnarray}
where each diagonal entry, $\textbfss{R}_{ij}$, represents the correlation coefficient between the $i$th and $j$th emission line parameter and $\textbfss{C}_{ij}$ is our full data covariance matrix defined as,
\begin{eqnarray}
\textbfss{C}_{ij} = \frac{1}{N-1}\sum^{N}_{i}(\textbfss{X}_{i} - \bmath{\mu}_{i})(\textbfss{X}_{j} - \bmath{\mu}_{j}).
\end{eqnarray}
Here, $\textbfss{X}_{i}$ is the data vector containing all $i$th emission line parameters from the full QSO sample and $\bmath{\mu}$ is its mean.

Relative to the other emission lines in our sample, the correlations with any \nv{} emission line parameters are moderate to weak. Primarily, \nv{} covaries strongest with the broad \lya{} component which makes sense under the assumptions of our model (fitting multiple Gaussian profiles). The extended wings of the broad \lya{} component can overlap with the \nv{} line component owing to their reasonably close separation ($\sim20$\AA). This is evident through the measured anti-correlation between the width of the broad \lya{} component and the width of the \nv{} line ($\rho=-0.32$). Note, there are a couple of other interesting correlations between the \nv{} line and individual line properties (e.g. amplitude or velocity offset). For example, between the velocity offsets of \nv{} and the narrow component of \civ{} and between the peak amplitude of the \nv{} and the \sioiv{} line profile. The former highlights the importance of including \nv{} into the damping wing procedure, as it implies that a strongly blue-shifted \civ{} narrow line component (as appears to be the case with these $z\gtrsim7$ QSOs) would result in a strongly blue-shifted \nv{} line either being close to or entering into our damping wing fitting region.

This implies that the reconstruction of the \nv{} emission is primarily dependent on the reconstruction of the \lya{} emission and is essentially independent on all other considered emission lines in our covariance matrix. Overall, this will result in larger uncertainties in the reconstructed profiles due to: (i) an increased number of parameters in the model and (ii) the weak dependency of \nv{} to all other lines producing a larger range of plausible profiles.

\section[]{Reanalysis of previous QSOs with \nv{} component} \label{sec:reanalysis}

As we have modified our reconstruction pipeline in this work to simultaneously predict the \nv{} emission line, it is prudent we perform a reanalysis of our previous work. In Section~\ref{sec:FitJ1120} and Section~\ref{sec:FitJ1342} we provide the updated reconstructed intrinsic QSO profiles along with the new constraints on the IGM neutral fraction for J1120 and J1342, respectively.

\subsection[]{ULASJ1120+0641} \label{sec:FitJ1120}

\begin{figure} 
	\begin{center}
	  \includegraphics[trim = 0.2cm 0.6cm 0cm 0.5cm, scale = 0.56]{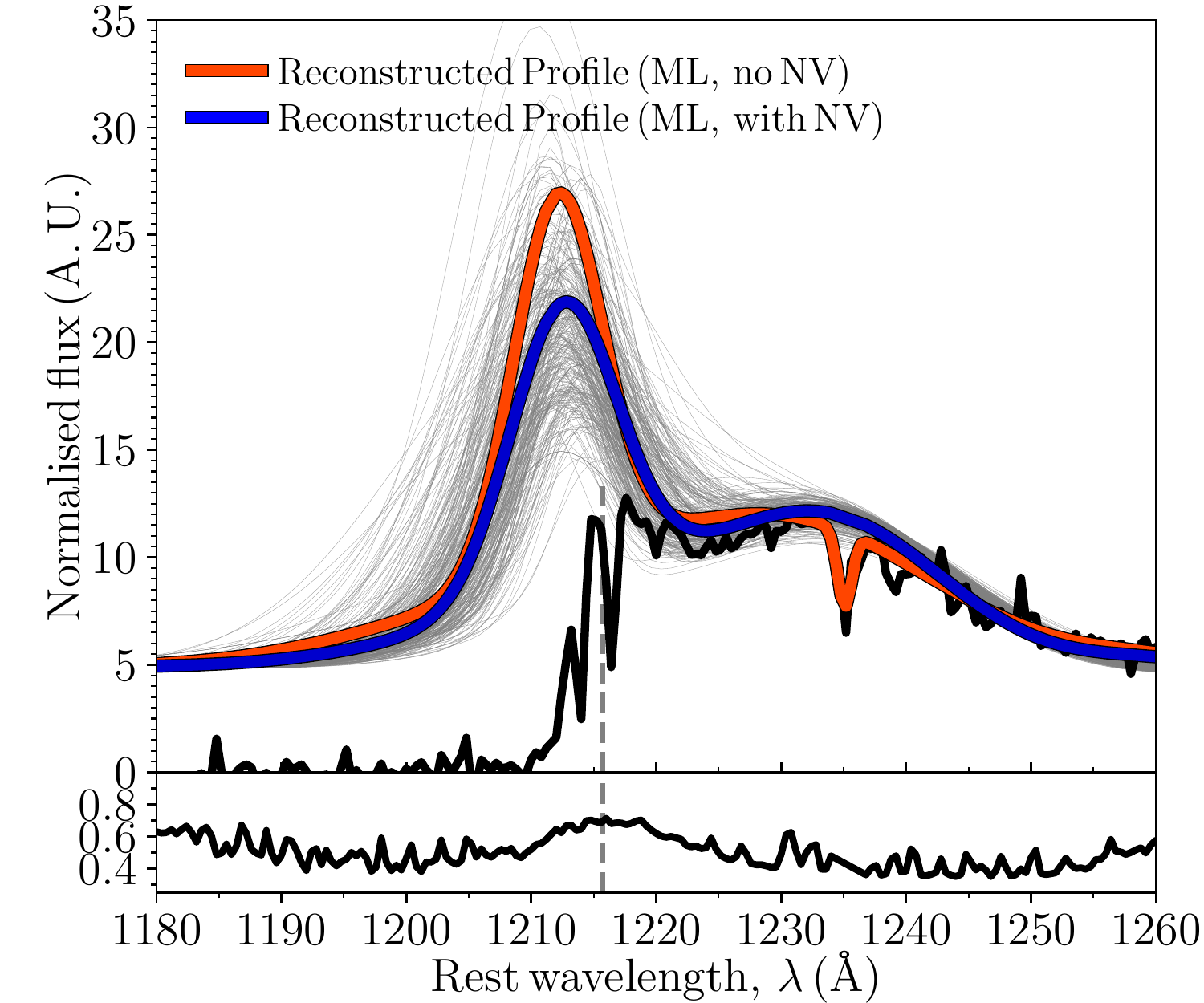}
	\end{center}
\caption[]{A reanalysis of the reconstructed maximum likelihood \lya{} emission line profile including a \nv{} line component for ULASJ1120+0641 ($z=7.08$). The red curve corresponds to the previous maximum likelihood reconstruction from \citet{Greig:2017a} whereas the blue curve is the new maximum likelihood reconstruction including the \nv{} component. Note previously, in \citet{Greig:2017a} features in the QSO spectrum were fit up to 1230\AA, thus \nv{} was fit rather than reconstructed. We also present a subsample of 300 \lya{} line profiles (thin grey curves) randomly drawn from the full posterior distribution of reconstructed profiles (including \nv{}). The black curve corresponds to the FIRE spectrum \citep{Simcoe:2012}, with the corresponding error spectrum shown in the sub-panel below. The vertical grey dashed line denotes rest-frame \lya{}.}
\label{fig:ULASJ1120}
\end{figure}

In Figure~\ref{fig:ULASJ1120} we present the updated maximum likelihood reconstruction for J1120 (blue curve) including the \nv{} line component relative to the original reconstruction \citep{Greig:2017} without \nv{} (red curve). The thin grey curves correspond to 300 randomly drawn profiles from the full posterior distribution of reconstructed profiles (including \nv{}). Firstly, it is important to note that in \citet{Greig:2017}, the reconstruction of the intrinsic profile used a flux prior over $\lambda=1230-1275$\AA. However, since \citet{Greig:2019} a flux prior has only been applied over $\lambda=1250-1275$\AA. As such, the original reconstruction for J1120 includes a \nv{} emission line as we were able to fit for it in combination with the \lya{} reconstruction.

Interestingly, the reconstructed \nv{} emission line closely resembles that of the fitted \nv{} emission line from \citep{Greig:2017}. This gives us confidence that the reconstruction pipeline works sufficiently well following the inclusion of \nv{} into the covariance matrix. The primary difference between the two reconstruction methods is the decrease in amplitude of the narrow \lya{} line component, along with a marginal broadening of the line. Further, the broad \lya{} line component is reduced in amplitude as indicated by the systematically lower flux amplitude for the full profile across $\lambda=1190-1230$\AA. This lower amplitude \lya{} profile somewhat reduces the difference in profile shapes between the covariance matrix and PCA pipelines.

For the updated reconstruction of J1120 we recover IGM neutral fraction constraints of $\bar{x}_{\hi{}} = 0.44\substack{+0.23 \\ -0.24}$ for the \intermediateHII{} EoR morphology. In \citet{Greig:2017}, we had previously recovered constraints of $\bar{x}_{\hi{}} = 0.46\substack{+0.21 \\ -0.21}$. Clearly, owing to the relatively similar amplitude and shape between the two reconstructions over the damping wing fitting region ($\lambda=1218-1230$\AA) we recover essentially the same constraints on the IGM neutral fraction. The main difference to note is that following the inclusion of \nv{} into the pipeline, the relative amplitude of the confidence intervals have increased owing to the increased scatter in reconstructed profiles due to the expanded covariance matrix. The slight decrease in the recovered IGM neutral fraction arises due to the decreased flux amplitude of the reconstructed profile between $\lambda=1220-1227$\AA. The decrease in the IGM neutral fraction however is minimised by the slight increase in predicted flux in the $\lambda=1218-1221$\AA\ region. The preference for a slightly wider narrow \lya{} line component will add some weight to damping wing profiles from a more neutral IGM.

\subsection[]{ULASJ1342+0928} \label{sec:FitJ1342}

\begin{figure} 
	\begin{center}
	  \includegraphics[trim = 0.2cm 0.6cm 0cm 0.5cm, scale = 0.56]{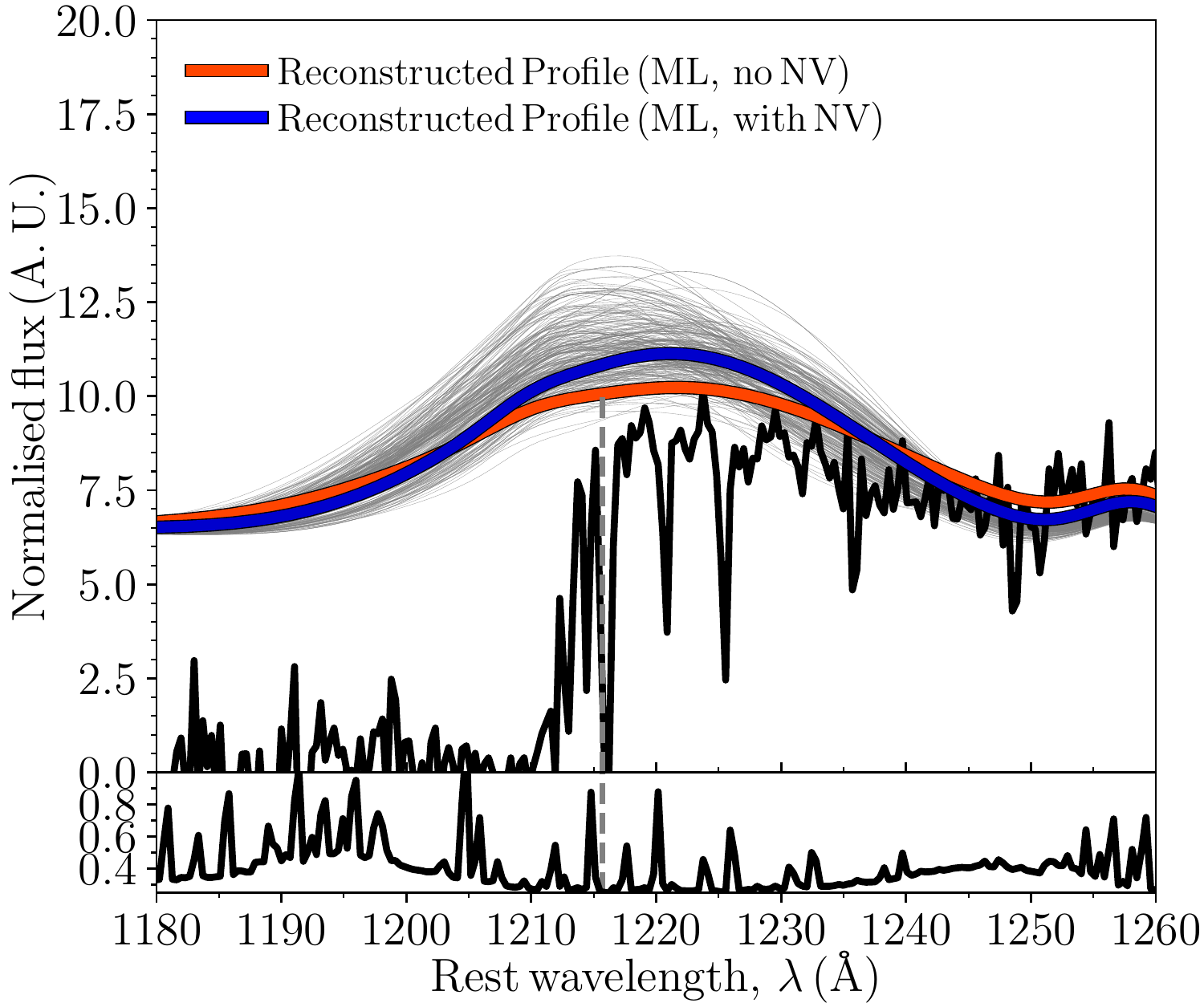}
	\end{center}
\caption[]{The same as Figure~\ref{fig:ULASJ1120}, except showing the reanalysis of ULASJ1342+0928 ($z=7.54$) originally presented in \citet{Greig:2019}. The black curve corresponds to the combined Magellan/FIRE and Gemini/GNIRS spectrum \citep{Banados:2018}, with the corresponding error spectrum shown in the sub-panel below. The vertical grey dashed line denotes rest-frame \lya{}.}
\label{fig:ULASJ1342}
\end{figure}

In Figure~\ref{fig:ULASJ1342} we present the updated maximum likelihood reconstruction for J1342 (blue curve) including the \nv{} line component relative to the original reconstruction \citep{Greig:2019} without \nv{} (red curve). Once again, the thin grey curves represent 300 randomly drawn profiles from the full posterior distribution of reconstructed profiles (including \nv{}).

Immediately evident is the increased amplitude of the predicted flux between $\lambda=1210-1230$\AA. This increase is primarily driven by a blue-shifted, broad \nv{} emission line. Equally, this broader \nv{} component results in a slightly narrower broad \lya{} component (due to the anti-correlation between the line widths discussed in Section~\ref{sec:Covariance_wNV}) which can then lead to an increased peak amplitude for the broad \lya{} component. Again, this reduces the difference between the maximum likelihood reconstruction profiles from the covariance matrix and PCA reconstruction pipeline highlighting the importance of including \nv{} into the reconstruction pipeline.

Performing our joint analysis for the IGM damping wing (\intermediateHII{} EoR morphology), we find $\bar{x}_{\hi{}} = 0.31\substack{+0.18 \\ -0.19}$ at 68 per cent confidence. In contrast, in \citet{Greig:2019} we recovered an IGM neutral fraction of $\bar{x}_{\hi{}} = 0.21\substack{+0.17 \\ -0.19}$. This increase in the recovered IGM neutral fraction is driven by the increased amplitude across the entire damping wing fitting region used in our analysis due to the addition of the \nv{} emission line component. Previously, for J1342 we had noted that the results from our covariance matrix approach was mildly in tension with the equivalent PCA analysis from \citet{Davies:2018}, who recovered constraints of  $\bar{x}_{\hi{}} = 0.60\substack{+0.20 \\ -0.23}$. While the results from the two pipelines are still notably different, the difference is now closer to $\sim1\sigma$ compared to the $\sim1.5\sigma$ reported previously.

In this work, we explored this difference in the recovered IGM neutral fraction a little further given that both pipelines now predict relatively similar reconstructed \lya{} profiles. To do this, we perform the same analysis as in Section~\ref{sec:JointFitting}, except we do not average over the full distribution of reconstructed profiles from our covariance matrix pipeline. That is, we only consider the maximum likelihood reconstruction profile while averaging over all synthetic damping wing profiles. In doing so, we recover $\bar{x}_{\hi{}} = 0.48\substack{+0.16 \\ -0.14}$. This implies that the full distribution of reconstruction profiles tends to sit on average lower in amplitude than the maximum likelihood profile or have reconstructed \lya{} profile shapes which prefer slightly more gradual damping wing profiles (i.e. lower amplitude neutral fractions) to match the observed spectrum. This goes some way towards explaining the difference between the two reconstruction pipelines. Further differences likely arise due to the shape of the reconstructed profiles blueward of \lya{}. The reconstructed profiles of the \citet{Davies:2018} PCA approach tend to be larger in amplitude blueward of \lya{} where they also fit for the damping wing profile (unlike only the redside in our work). Thus, having higher amplitude reconstructed profiles at these blueward wavelengths would prefer higher IGM neutral fraction values to be able to suppress the reconstructed profiles to match the observed flux.

\section[]{MCMC Fitting of the QSOs} \label{sec:MCMCQSOs}

\begin{figure*} 
	\begin{center}
	  \includegraphics[trim = 0.5cm 0.6cm 0cm 0.5cm, scale = 0.475]{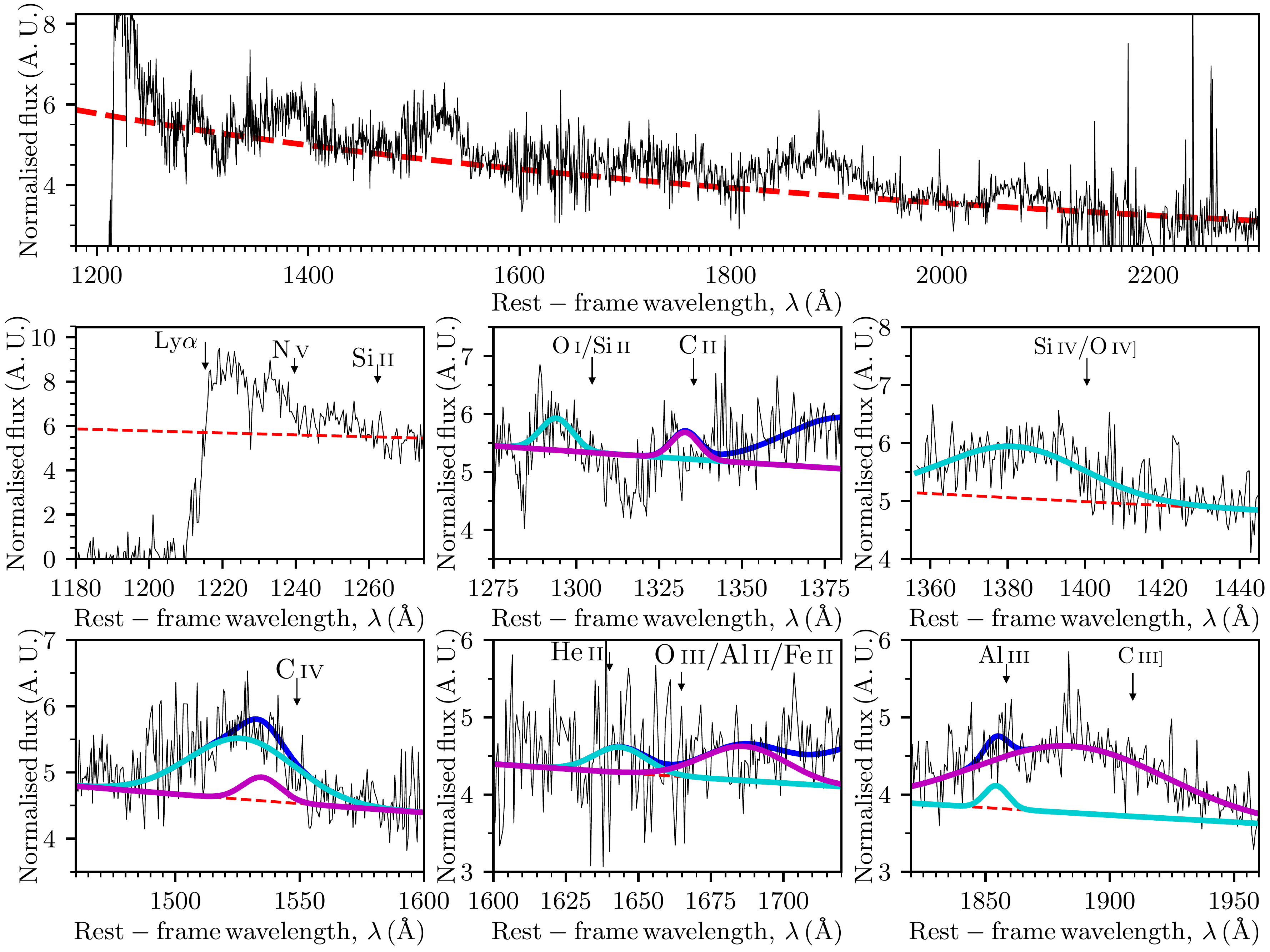}
	\end{center}
\caption[]{A zoom-in highlighting the MCMC fitting procedure from \citep{Greig:2017a} applied to the combined GNIRS and NIRES spectrum of J1007 \citep{Yang:2020}. The QSO continuum flux is normalised at 1450\AA\ rest frame ($f_{\lambda} \propto \left(\frac{\lambda}{\rm 1450\AA}\right)^{\alpha}$; 1 a.u. = 10$^{-18}$ cm$^{-2}$ s$^{-1}$ \AA$^{-1}$). Top panel: a single power-law continuum (red dashed curve) is fit to the QSO spectrum. Middle left: the attenuated \lya{} profile. Middle centre: the low ionisation lines, O {\scriptsize I}/Si {\scriptsize II]} (cyan), and \cii{} (magenta). Middle right: the blended \sioiv{} line complex fit with a single component Gaussian. Bottom left: the \civ{} line fit with a double component Gaussian. Bottom centre: Low ionisation lines, \heii{} and \oiii{}/Al\,{\scriptsize II}/Fe\,{\scriptsize II}. Bottom right: single component Gaussians to describe the \ciii{} line (magenta) and Al {\scriptsize III} (cyan) emission lines.}
\label{fig:FitJ1007}
\end{figure*}

\begin{figure*} 
	\begin{center}
	  \includegraphics[trim = 0.5cm 0.6cm 0cm 0.5cm, scale = 0.475]{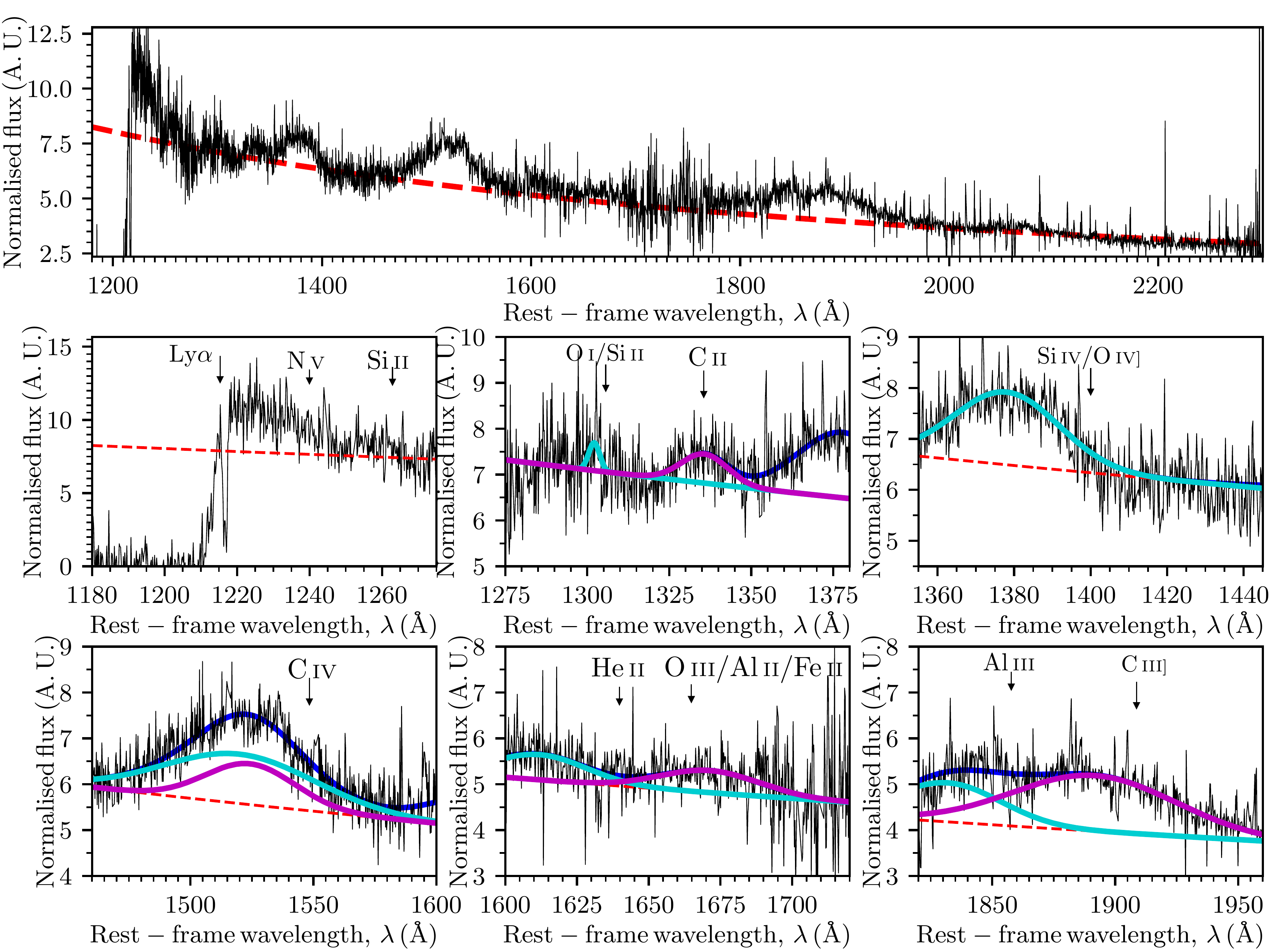}
	\end{center}
\caption[]{Same as Figure~\ref{fig:FitJ1007}, except applied to the Gemini/GMOS + Keck/NIRES spectrum of J0252 \citep{Wang:2020}.}
\label{fig:FitJ0252}
\end{figure*}

In order to perform our covariance matrix reconstruction pipeline, we perform an MCMC fit to the observed spectrum over the range $\lambda =1275-2300$\AA\ (see Section~\ref{sec:Reconstruction} for further details). In Figure~\ref{fig:FitJ1007}, we provide the MCMC fit to J1007. In the top panel, we show the two component power-law QSO continuum (red dashed curve), while in the remaining panels we provide zoomed-in panels centred on all the emission lines that we fit in our MCMC pipeline. In Figure~\ref{fig:FitJ0252}, we provide the equivalent MCMC fit to J0252.

\end{document}